\documentclass[pre,aps,twocolumn,showpacs,floatfix]{revtex4}
\usepackage{graphicx}
\usepackage{epsfig}
\usepackage{epstopdf}
\usepackage{amsmath}
\usepackage{amssymb}

\begin{document}

\title{Energetics of discrete selectivity bands and mutation-induced transitions \\in the calcium-sodium ion channels family}

\author{I.~Kaufman$^{1}$, D.~G.~Luchinsky$^{1,2}$, R.~Tindjong$^1$, \\
           P.V.E.~McClintock$^1$, R.S.~Eisenberg$^3$}

\affiliation{$^1$Department of Physics, Lancaster University, Lancaster LA1 4YB, UK}
\email{p.v.e.mcclintock@lancaster.ac.uk}

\affiliation{$^2$Mission Critical Technologies Inc., 2041 Rosecrans Ave. Suite 225 El Segundo, CA 90245, USA}

\affiliation{$^3$Department of Molecular Biophysics and Physiology, Rush Medical College, 1750 West Harrison, Chicago,
IL 60612, USA}

\date{\today}

\begin{abstract}
We use Brownian dynamics simulations to study the ionic conduction and valence selectivity of a generic electrostatic model of a biological ion channel as functions of the fixed charge $Q_f$ at its selectivity filter. We are thus able to reconcile the discrete calcium conduction bands 
recently revealed in our BD simulations, M0 ($Q_f=1e$), M1 (3$e$), M2 (5$e$), with a set of sodium conduction bands L0 (0.5$e$), L1 (1.5$e$), thereby obtaining a completed pattern of conduction and selectivity bands {\it vs.} $Q_f$ for the sodium-calcium channels family.  An increase of $Q_f$ leads to an increase of calcium selectivity: L0 (sodium selective, non-blocking channel) $\rightarrow$ M0 (non-selective channel) $\rightarrow$ L1 (sodium selective channel with divalent block) $\rightarrow$ M1 (calcium selective channel exhibiting the anomalous mole fraction effect). We create a consistent identification scheme where the L0 band is putatively identified with the eukaryotic (DEKA) sodium channel 
The scheme created is able to account for the experimentally observed mutation-induced  transformations between non-selective channels, sodium-selective channels, and calcium-selective channels, which we interpret as transitions between different rows of the identification table.
By considering the potential energy changes during permeation, we show explicitly that the multi-ion conduction bands of calcium and sodium channels arise as the result of resonant barrier-less conduction.
The pattern of periodic conduction bands is explained on the basis of sequential neutralisation taking account of self-energy, as $Q_f(z,i)=ze(1/2+i)$, where $i$ is the order of the band and $z$ is the valence of the ion. Our results confirm the crucial influence of electrostatic interactions on conduction and on the Ca$^{2+}$/Na$^+$ valence selectivity of calcium and sodium ion channels. The model and results could be also applicable to biomimetic nanopores with charged walls.
\end{abstract}

\pacs{
87.16.Vy, 
41.20.Cv, 
05.40.-a, 
87.10.Mn  
}


\maketitle
\section{Introduction}\label{Sec:introduction}
At the molecular level, an understanding of living systems requires the application of physics and this is particularly true in the case of biological ion channels. Here, we study the physics of a simple electrostatic model to investigate the operation of voltage-gated calcium and sodium ion channels. Their importance stems from
their essential roles in controlling muscle contraction, neurotransmitter secretion, gene regulation and the transmission of action potentials. The effective function of calcium channels is based on their high selectivity for divalent calcium ions Ca$^{2+}$ over monovalent sodium ions Na$^+$. They exhibit the anomalous mole fraction effect (AMFE), an effective blockade of Na$^+$ permeation by small concentrations of Ca$^{2+}$, combined with measurable Ca$^{2+}$ currents in the pA range \cite{Hille:01,Sather:03}. Sodium channels have very similar structure but demonstrate the opposite kind of selectivity, favouring  Na$^+$ over Ca$^{2+}$.

The selectivity of calcium and sodium channels is defined by a narrow selectivity filter with a strong binding site. The latter is formed of protein residues with a net negative charge $Q_f$ whose magnitude depends on the particular residues \footnote{The protein residues are amino acids, of which aspartate (D) and glutamate (E) have negatively charged side chains. Others that we mention here are lysine (K), which has a positively charged side chain, as well as alanine (A), leucine (L), tryptophan (W) and serine (S).} that are present. The L-type calcium channel has a highly-conserved EEEE locus with four glutamates \cite{Yang:93}, whereas the RyR channel has a DDDD locus \cite{Gillespie:08}. The DDDD locus was also found  in the TRPV6 transient receptor potential channel, which is highly calcium-selective but generally very different from the RyR channel \cite{Owsianik:06}.

Although sodium and calcium channels have similar structures, they have different selectivity filter loci (and therefore different $Q_f$), and have different lengths and radii \cite{Vora:05,Payandeh:11,Csanyi:12}. The eukaryotic sodium channel has two charged rings at or near the selectivity filter: an inner DEKA ring with a nominal $Q_f = 1e$ and an outer EEDD ring with nominal $Q_f=4e$ \cite{Vora:05,Csanyi:12, Catteral:12} where $e=-1.6 \times 10^{-19}$\,C is the electronic charge. Bacterial sodium channels can have rather different selectivity filter loci and represent  L-type-like EEEE locus in the NaChBac and recently-studied  NavAb channel \cite{Payandeh:11,Payandeh:12}.

Experimental studies of mutations in the protein side chains \cite{Heinemann:92, Schlief:96, Tang:93, Shaya:11, Burgess:99, Koch:00, Yue:02, Miedema:04, Senatore:13},  and model simulations \cite{Boda:07, Boda:08, Csanyi:12}, show that the value of $Q_f$ is a crucial factor in determining the Ca$^{2+}$ {\it vs.} Na$^+$ selectivity of calcium and sodium channels. Usually, mutations that influence $Q_f$ also destroy the calcium channel's selectivity, and hence physiological functionality, leading to ``channelopathies" \cite{Burgess:99, Hubner:02}. However, an appropriate point mutation of the DEKA sodium channel ($Q_f \approx 1e$) converts it into a calcium-selective channel with a DEEA locus and $Q_f \approx 3e$ \cite{Heinemann:92}. The essentially non-selective bacterial OmpF porin ($Q_f \approx 1e$) can be converted into a Ca$^{2+}$-selective channel by the introduction of two additional glutamates in the constriction zone; the resultant mutant contains a DEEE-locus ($Q_f \approx 4e$) and exhibits an Na$^+$ current with a strongly increased sensitivity to 1\,mM Ca$^{2+}$ \cite{Miedema:04}. Simulations \cite{Csanyi:12} have indicated growth of Ca$^{2+}$ {\it vs.} Na$^+$ selectivity as $Q_f$ increases from $1e$ to $4e$.

The mechanisms of Ca$^{2+}$/Na$^+$ selectivity underlying these transformations have remained unclear, as has also the unambiguous identification of the ``charge -- selectivity type" relationship.

Multi-ion  knock-on barrier-less  conductivity \footnote{Note that, strictly, it is low-barrier conduction: the potential barriers are still present albeit greatly reduced in size, as discussed in Sec.\ \ref{Sec:results}. For convenience, however, we will follow the convention of referring to ``barrier-less conduction.''} is assumed to be one of the main mechanisms of permeation and selectivity for the potassium \cite{Hodgkin:55, Roux:04} and calcium \cite{Hess:84, Armstrong:91} channels, and is inferred to be a general mechanism of selectivity \cite{Yesylevskyy:05}. Barrier-less knock-on conductivity can also be described as a limiting case of long-range ion-ion correlations \cite{Luchinsky:09b,Tindjong:12a}.

Generic electrostatic models describe an ion channel as a cylindrical water-filled hole in a charged protein in the cell membrane \cite{Levitt:78,Laio:99,Cheng:05}. They usually assume single-file motion of the permeating ions and can reproduce significant features related to the conductivity and selectivity \cite{Nadler:03,Corry:05, Zhang:05, Kharkyanen:10, Eisenberg:13a}. Thus a single model with almost unchanging parameters can account for the valence selectivity features of both sodium and calcium channels (reviewed in \cite{Eisenberg:12b, Eisenberg:13c}). An analytic treatment of such a model \cite{Zhang:05, Kamenev:06, Zhang:06} showed that transport of Ca$^{2+}$ ions through a negatively-doped channel 
exhibited several 
ion-exchange phase transitions as functions of bulk concentration and $Q_f$, with a near-zero transport barrier at the transition points \cite{Zhang:06}. 
Brownian dynamics (BD) simulations of 
L-type calcium channel revealed a narrow peak in Ca$^{2+}$ conductance near $Q_f$ =3.2$e$ \cite{Corry:01}. Discrete multi-ion conduction peaks were predicted in \cite{Kharkyanen:10}. The possibility that channel conduction might be a discontinuous function of channel parameters, with pass bands and stop bands
having been discussed at length in the speculations of one of us, long ago \cite{Eisenberg:96}.

We have recently used parametric Brownian dynamics (BD) simulations of ionic currents for different $Q_f$ in a generic model of calcium channels to show that the Ca$^{2+}$ conduction and Ca$^{2+}$/Na$^+$ valence selectivity form a regular pattern of narrow conduction/selectivity bands as a function of $Q_f$, separated by regions of non-conduction. These discrete bands relate to saturated, self-sustained Ca$^{2+}$ conductivity with different numbers of ions involved in the conduction; they correspond to the phase transitions obtained analytically in \cite{Zhang:06} and are consistent with earlier results \cite{Corry:01, Kharkyanen:10}. We have associated the underlying mechanism with multi-ion barrier-less conductivity, identified the calcium selective bands seen in the simulations with known calcium channels, and inferred that the band structure could explain the results of mutant studies \cite{Kaufman:13a, Eisenberg:13b}.

We also investigate the energetics of the generic electrostatic model and consider potential energy profiles along optimal multi-ion stochastic  trajectories \cite{Dykman:92a,Kharkyanen:10} to show that the calcium and sodium conduction and selectivity bands are based on the barrier-less conduction mechanism \cite{Kaufman:13c}.

In this paper we complete the ordered sequence of Ca$^{2+}$/Na$^{+}$ conductivity and selectivity bands {\it vs.} surface charge $Q_f$ for the sodium-calcium channels family, initiated in \cite{Kaufman:13a}. We add an analysis of sodium bands and construct an identification table to explain and classify numerous mutation-induced transformations of Ca$^{2+}$/Na$^+$ selectivity in the calcium-sodium channels family.

We start by summarising in Sec.\ \ref{Sec:genericmodel} the main features of the generic model. In Sec.\ \ref{Sec:pattern} we describe the ordered sequence of selectivity types for the sodium-calcium family of channels based on BD simulations of the model \cite{Kaufman:13a}. We relate these data to  real ion channels in Sec.\ \ref{Sec:identification} and to mutation-induced transformations between them in \ref{Sec:mutations}. In Secs.\ \ref{Sec:energetics1ion} and \ref{Sec:energetics2ions} we work out the energetics of permeation and show how the observed bands correspond to optimal conditions (minimal energy barrier) for one-ion and two-ion processes respectively.
In Sec.\ \ref{Sec:neutral} we discuss the patterns of bands for different ions that result from a neutralisation approach. Finally, in Sec.\ \ref{Sec:Conclusions} we summarise and draw conclusions.

\section{A generic electrostatic model of calcium channels}\label{Sec:genericmodel}
\begin{figure}[t]
\begin{center}
\includegraphics[width=1.0\linewidth]{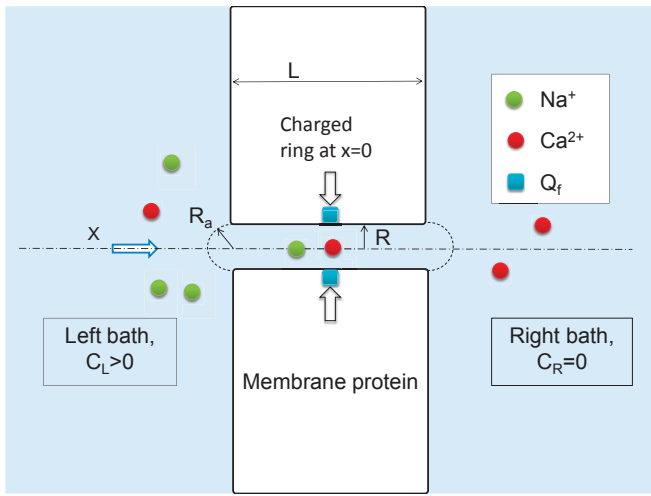}
\end{center}
\caption{(Color online) Computational domain for a generic model of the calcium ion channel (reworked from \cite{Kaufman:13a}). Its selectivity filter is treated as an axisymmetric, water-filled, cylindrical hole of radius $R=3$\,\AA ~and length $L=12-16$\,\AA ~through the protein hub in the cellular membrane. The $x$-axis is coincident with the channel axis and $x=0$ in the middle of the channel. There is a centrally-placed, uniformly-charged, rigid ring of negative charge $Q_f=0-6.5e$ embedded in the wall at $R_Q=R$. The left-hand bath, modeling the extracellular space, contains non-zero concentrations of Ca$^{2+}$ or Na$^+$ ions. These are injected on the axis at the Smoluchowski diffusion rate at a distance $R_a=R$ outside the left-hand entrance. The domain length $L_d$=100\,\AA, the domain radius $R_d=100$\r{A}, the grid size $h=0.5$\,\AA, and a potential difference of 0--75\,mV is applied between the left and right domain boundaries.} \label{fig:channel}
\end{figure}

\subsection {Geometry and general features of the model}
Fig.\ \ref{fig:channel} shows the generic, self-consistent, electrostatic model of a calcium/sodium channel. We focus exclusively on its selectivity filter, which we consider as a negatively-charged, axisymmetric,  water-filled, cylindrical pore of radius $R=3$\,\AA ~and length $L=12-16$\,\AA ~through the protein hub in the cellular membrane  The $x$-axis is coincident with the channel axis and $x=0$ in the center of channel. There is a centrally-placed, uniformly-charged, rigid ring of negative charge $Q_f=0-6.5e$ embedded in the wall at $R_Q=R$. The left-hand bath, modeling the extracellular space, contains non-zero concentrations of Ca$^{2+}$ and/or Na$^+$ ions. In the simulation, these are injected on the axis at the Smoluchowski diffusion rate at a distance $R_a$ from the channel entrance. The domain length $L_d$=100\,\AA, the domain radius $R_d=100$\r{A}, the grid size $h=0.5$\,\AA, and a potential difference in the range 0--75\,mV is applied between the left and right domain boundaries.

This generic model of a calcium ion channel is similar to that used previously \cite{Nonner:00,  Corry:01, Giri:11}.  Details of the model have already been presented and its validity and limitations discussed \cite{Kaufman:13a}, but for completeness we now summarise  and give some additional details and discussion.

The minimum possible radius $R$ of the selectivity filter of an L-type calcium channel has been determined as being $R=2.8$\,\AA. We use the value of $R=3.0$\,\AA. The mobile sodium and calcium ions are described as charged spheres of radius $R_i\approx 1$\,\AA~ (matching both ions), with diffusion coefficients of $D_{Na}=1.17 \times 10^{-9}$ m$^2$/s and $D_{Ca}=0.79 \times 10^{-9}$ m$^2$/s, respectively. In what follows we assume an asymmetrical ionic concentration: $C_L>0$ on the left, and $C_R=0$ on the right, corresponding to the physiological conditions in calcium and sodium channels.

We take both the water and the protein to be homogeneous continua with dielectric constants $\varepsilon_w=80$ and $\varepsilon_p=2$, respectively, together with an implicit model of ion hydration (the validity of which is discussed elsewhere). We approximate $\varepsilon_w$ and $D$ as equal to their bulk values throughout the whole computational domain, including the selectivity filter, a choice that avoids the use of arbitrary fitting parameters.

The importance of self-consistent calculations cannot be overstated. If calculations are not self-consistent, then the potential does not take proper account of all the charges that are present. Thus, some of the potential then has a mysterious nonphysical origin. In the real world, and in experiments, conditions and concentrations change. Consistent calculations determine and follow the potential that results from these changes \cite{Eisenberg:13c}.

Our simulation scheme could be described as a self-consistent numerical solution of Poisson's electrostatic equation coupled with the Langevin stochastic equation for the moving ions.

\subsection {Self-consistent electrostatics for generic ion channel geometry}

The electrostatic potential $U$ for an ion and the potential gradients were derived by numerical solution of Poisson's equation within the computational domain shown in Fig.\ \ref{fig:channel}:
\begin{equation}
-\nabla (\varepsilon_0 \varepsilon \nabla U) = \rho_0+\sum_i e z_i n_i
\label{eq:poisson}
\end{equation}
where $\varepsilon_0$ is the dielectric permittivity of vacuum, $\varepsilon$ is the dielectric permittivity of the medium (water or protein), $\rho_0$ is the density of fixed charge, $z_i$ is the charge number (valence), and $n_i$ is the number density of moving ions. We used a 2D/3D axisymmetric finite-volume Poisson solver with a staggered grid, specially designed to accommodate the large permittivity mismatch \cite{Oevermann:06, Kaufman:09} at the water-protein interface.

We utilized field linearity and the superposition rule to speed-up the run-time calculations. The potential $U$ and electrostatic field $E$ were pre-calculated for all axial ion positions on the grid and saved in look-up tables that were using during run-time for quick recovery of the relevant $U$ and $E$ values \cite{Hoyles:98}. In doing so, full account was taken of the static charge, interactions, and self-energy contributions.

\begin{figure*}[t]
\includegraphics[width=1.0\linewidth]{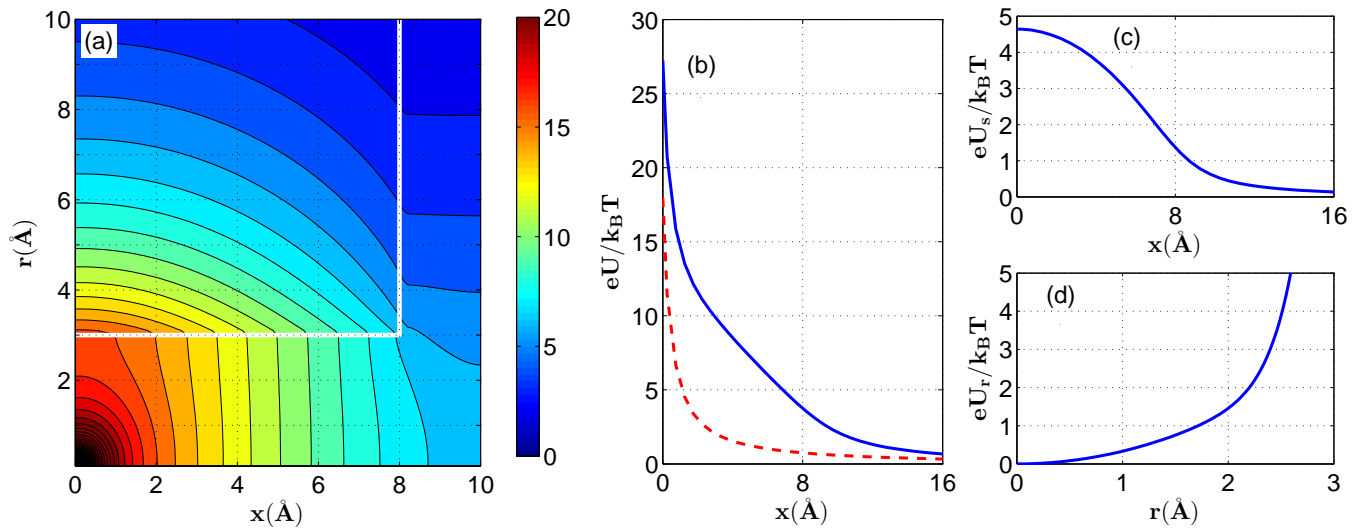}
\caption{(Color online)  Polarisation effects in the generic model of ion channel. (a) Electrostatic potential map $U(x,r)$ for zero membrane potential and centered monovalent cation inside the selectivity filter, whose spatial limits are indicated y the white lines. Colorbar is in units  of $k_BT/e$, contour step is 1 in the same units. The equipotential lines are almost equally-spaced and almost perpendicular to the $x$-axis, illustrating the quasi-1D behavior of the electrostatic field. (b) Electrostatic amplification of the electrostatic field inside the channel: the potential $U$ of a monovalent ion within the channel (blue, full curve) significantly exceeds the corresponding potential in bulk water (red, dashed) $U_0$ due to induced polarisation charge that appears at the water/protein interface.  (c) The axial self-energy potential $U_s$ accounts for the dielectric contribution to the hydration barrier. (d) the radial self-energy $U_r$ provides a stable point in the center of the channel cross-section, at $r=0$.} \label{fig:electro}
\end{figure*}

Self-consistent electrostatics within the narrow, water-filled, channel in the protein differs significantly from bulk electrostatics, even when the dielectric constant of the water inside the channel is taken to be the same as in the bulk. The huge gradient  between $\varepsilon_w=80$ and $\varepsilon_p=2$, and the specific channel geometry, lead to a number of effects that are crucial for ion permeation through the channel \cite{Edwards:02, Cheng:05}, as illustrated in Fig. \ref{fig:electro}.

Fig. \ref{fig:electro}(a) shows that the boundary conditions at the water/protein interface almost eliminate the radial component $E_r$ of the elecrostatic field in comparison with the axial component $E_x$. Thus $E_r \ll E_x$. It is this condition that results in a quasi-1D axial behaviour of the electrostatic field (constant $E_x$ due to a linear variation in $U$ between the charges), and hence in preferentially axial motion of ions inside the channel \cite{Cheng:05, Zhang:05}, which behave like a one-dimensional Coulomb gas \cite{Kamenev:06}.

Fig. \ref{fig:electro}(b) demonstrates electrostatic amplification of the electric field inside the narrow channel due to partial de-screening of the electrostatic field and the appearance of polarisation charges at the water/protein boundary, which is what results in the quasi-1D field behavior \cite{Laio:99, Zhang:05, Luchinsky:09d}.

Fig. \ref{fig:electro}(c) illustrates a remarkable feature of the channel geometry that strongly influences permeation: the high self-energy barrier $U_s$ corresponding to the dielectric boundary force. It amounts to an electrostatic contribution to the free energy barrier and thus adds to the dehydration barrier \cite{Cheng:05, Berti:12a}. This barrier prevents any ion from entering an empty uncharged channel: an ion in bulk is repelled from the boundary with the protein. The electrostatic component of the barrier is independent of the ionic radius \cite{Nadler:03} and, as we will see below, it can help to account for valence selectivity. Incorporating of more advanced hydration models \cite{Zwolak:09} could provide alike selectivity as well.

Fig. \ref{fig:electro}(d) shows the radial self-energy potential profile $U_r$, representing a  potential well centred on the channel axis. Hence an ion inside the channel experiences a radial force towards the axis. The existence of this force helps to justify the conventional approximation of strictly axial single-file movement of ions inside the narrow channel \cite{Cheng:05, Zhang:05}.

The electrostatics of an empty ion channel prohibits the entry of any negatively charged (e.g.\ chloride) ion due to combined influences of the dielectric boundary force and the repulsion of the fixed negative charge \cite{Edwards:02, Nadler:03}. For this reason we take no account of counterions in the electrostatics and BD simulations. When the channel is occupied by cations, however, it becomes easier for anions to enter.

Consequently, we use a 1D dynamical model to simulate the axial single-file movement of cations (only) inside the selectivity filter and in its close vicinity.  Some additional discussion of these approximations is provided in  Sec.\ \ref{Sec:genericmodel}D.

\subsection {Brownian dynamics simulation of ionic current} \label{Sec:BD}
The BD simulations were based on numerical solution of the 1D over-damped time-discretized Langevin equation for the $i$-th ion:
\begin{equation}
\frac {d x}{dt} = -D z \left(\frac {\partial U}{\partial x}\right) + \sqrt{2D }  \xi (t)
\label{eq:langevin}
\end{equation}
where $D$ is the ionic diffusion coefficient, $\xi(t)$ is normalized white noise, $z$ is the valence of the ion, and the potential $U(x)$ is given in $(k_BT/e)$ units where $T$ is the temperature and $k_B$ is Boltzmann's constant. Numerical solution of (\ref{eq:langevin}) was implemented with the Euler forward scheme. Poisson's equation (\ref{eq:poisson}) is solved self-consistently at each simulation step as described above.

We use an ion injection scheme that allows us to avoid wasteful and heavy-duty simulation of ionic movements in the bulk liquid. The model includes a hemisphere of radius $R_a=R$ at each entrance representing the boundaries between the channel vicinity and the baths. The arrival rate $j_{\rm arr}$  is connected to the bulk concentration $C$ through the Smoluchowski diffusion rate: $j_{\rm arr}=2\pi D R_a C$ \cite{Nadler:01,Nadler:03,Luchinsky:09d}.

We model the ions as ``transparent'' in the sense that they can, in principle, pass each other. However, this is unlikely to happen. This simplification is based on the momentum conservation law for alike ions and also because quasi-1D electrostatic repulsion amplified by the narrow channel is strong enough to effectively prevent ions (whether alike or different) from coming close. Thus the assumption of single-file behavior is a good approximation (see details in the next section).

The motion of each injected ion is simulated in accordance with (\ref {eq:poisson}) until it reaches a domain boundary, where it is assumed to be absorbed.  The simulation continues until a chosen simulation time has been reached. The ionic current $J$ is calculated as the averaged difference between the numbers of similar ions passing the central cross-section of the channel per second in the forward and reverse directions \cite{Yesylevskyy:05}.

Quantities measured during the simulations include the sodium $J_{Na}$ and calcium $J_{Ca}$ ion currents, the partial ionic occupancy profiles $\rho(x)$ along $x$ for different concentrations, and the partial $P_{Na}$ and $P_{Ca}$ occupancies, in each case as functions of the respective concentrations of calcium $[Ca]$ or sodium $[Na]$.

The BD simulations of ion current $J$ and occupancy $P$ were performed separately for CaCl$_2$ and NaCl solutions, and also for a mixed-salt configuration, with concentrations $[Na]=30$mM and $20 \mu  $M$ \le [Ca] \le 80$mM. The value of $Q_f$ was varied  within the range 0--6.5$e$ in order to cover the known variants of sodium and calcium channels (\cite{Csanyi:12}).

\subsection {Validity and limitations of generic model} \label{Sec:validity}

Our reduced model obviously represents a considerable simplification of the actual electrostatics and dynamics of moving ions and water molecules 
within the narrow selectivity filter \cite{Tieleman:01,Nelissen:07}. We now discuss briefly the main simplifications limiting its validity: the use of continuum electrostatics; the use of BD; and the assumption of 1-D (i.e.\ single-file) movement of ions inside the selectivity filter.

The validity of both the electrostatics and the dynamics depends on the degree of dehydration of the ion inside the channel, so it can be defined roughly by the relationship between the channel radius $R$ and the radius of the ion’s first hydration shell $R_h$. Continuum electrostatics and dynamics generally fail when $R_h > R$, but still can be applied for $R_h \approx R$ provided that one uses effective values of $\varepsilon_w$ and the diffusion coefficients $D_{Na}$, $D_{Ca}$ that are all dependent on $R$ \cite{Laio:99}.

We estimate $R_h \approx 3.5$\AA ~for Na$^+$ and Ca$^{2+}$ ions, so that the calcium channel of $R \approx 3 $\AA ~\cite{Sather:03} does provide some room for Na$^+$ and Ca$^{2+}$ ions  to carry water molecules. Both ions are still partially hydrated, therefore, and the continuum approximation with effective values can be used inside the selectivity filter. It is shown in \cite{Laio:99} that the effective value of $\varepsilon_w$ saturates to its bulk value $\varepsilon_w$=80 for $R \approx 3.5 $\AA ~(roughly corresponding to $R_h$) and is still close to it ($\varepsilon_w$≈70) for $R=3$\AA. This allows us to use the bulk value for $\varepsilon_w$. The effective values of the ionic diffusion coefficients also decrease significantly with decreasing $R$ compared to their bulk values, and are estimated as $D\approx 0.25 D_{\rm bulk}$ for $R=3 $\AA ~\cite{Tieleman:01}.  As a result we can assume that in our channel model both ions move along the axis with almost unbroken first hydration shells.

We therefore use the standard bulk values of $\varepsilon_w=80$, $\varepsilon_w=2$ and $D$ as effective values throughout the whole computational domain, including the selectivity filter, a choice that avoids the use of additional fitting parameters.

The single-file condition can become a significant restriction if multiple ions are occupying the channel. In our model, however, single-file movement appears not as an {\it a priori} assumption but as the outcome of the Langevin dynamics of movement under electrostatic forces in a confined environment.  Thus single-filing of ions within the selectivity filter of the calcium channel is provided, not by direct geometrical restrictions, but by the combined effect of the above-mentioned self-repulsion from the channel wall together with strong mutual electrostatic repulsion between the moving ions. The minimum spatial separation between their centres needed for Ca$^{2+}$ or Na$^+$ ions  to pass each other is $d_{min}=2$\AA, with a maximum possible passing distance of $d_{max}=4$\AA ~within the 6\AA ~diameter channel. Even for monovalent ions $d_{max}$ is shorter than the Bjerrum length $l_B$ defined as the average distance for the thermal separation of charged ions ($U(l_B)=k_BT$): for water $l_B \approx 7 \AA$ is almost  $2 d_{max}$ and thus the probability of ions passing each other is low, especially taking into account the additional force due to self-repulsion from the wall.
In the most significant Ca$^{2+}$-Na$^+$ ``blockade" we can estimate the total energy barrier impeding leakage  as being $\approx 6k_BT$, which is high enough to justify our assumption of single-file movement.

As noted above in Sec.\ \ref{Sec:genericmodel}B, a shortcoming of the model is that, although it treats sodium and calcium ions explicitly, it fails to include their counter-ions (chloride), either explicitly or implicitly, in the Poisson continuum treatment and BD simulations. We justify this simplification through consideration of the combined effects of the self-potential barrier and repulsion by the fixed charge, both of which tend to prevent counter-ions from entering the empty selectivity filter. The situation is different for an occupied channel, however, and in certain cases chloride ions will then be able to enter \cite {Zhang:06}. Nonetheless,  experiments and simulations show that the concentration of chloride ions inside the channel is in practice reasonably small \cite{Boda:13}.

An important criterion for the applicability of a channel model is its ability to reproduce AMFE and, in particular, low-offset calcium blockade of the sodium current for the L-type (EEEE) calcium channel. In real experiments, blockade can be seen for $[Ca]_{50}<1\mu$M \cite{Sather:03}, and the same offset has been obtained in Monte-Carlo simulations \cite{Boda:08} and (indirectly) in BD simulations \cite{Corry:01}. Our BD simulations yield a blockade offset of about $[Ca]_{50}=40 \mu$M, which may be regarded as reasonable given the simplifications of the model.

The DEKA sodium channels, and mutants, might seem to stretch the generic model in that the ring of fixed charge is in reality fragmented around the pore (rather than being continuous), and is asymmetrical.  However, it is known that the axial field of a fragmented ring is exactly the same as for a continuous ring due to the high symmetry of the protein segments \cite{Sather:03}. Asymmetry of the DEKA ring is found to be significant for selectivity between alike ions \cite{Corry:13}; here, however, we study valence selectivity which depends mainly on the {\it total} charge at the selectivity filter \cite{Boda:08, Miedema:04}.

Generally, simulations based on simplified models \cite{Corry:01,Boda:08,Kaufman:13a} reproduce reasonably well the signatures of calcium channels, such as their AMFE \cite{Sather:03}. Despite their simplified nature, models of this sort can account quantitatively for the detailed properties of the RyR channel  and have enabled the prediction of complex current--voltage relationships in advance of the corresponding experiments, with errors of less than 10\% \cite{Gillespie:08}.


To summarize, the model is generic in the sense that it is just based on electrostatics and on the fundamental physical properties of channels of simplified geometrical shape. It takes no account of the detailed structure of the proteins or residues, and it treats water and protein as continuum dielectrics with their bulk dielectric constants. It could equally well be applied to e.g.\ TPRV channels \cite{Owsianik:06} and, because there is nothing inherently ``biological'' about it, the model should also be applicable to biomimetic nanotubes \cite{GarciaFandino:12,Miedema:12} and other artificial pores.

\section{Results and discussion}\label{Sec:results}

\subsection{The pattern of calcium and sodium conduction and selectivity bands}\label{Sec:pattern}
\begin{figure}[t]
\includegraphics[width=1.0\linewidth]{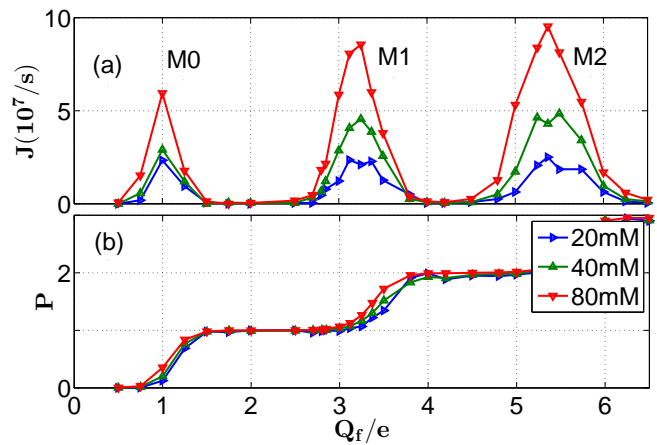}
\caption{(Color online)  BD simulations showing calcium conduction and occupancy bands of the generic ion channel model (partly reworked from \cite{Kaufman:13a}). (a) Plots of the calcium ionic current $J$ as a function of the fixed charge $Q_f$ at the selectivity filter for pure calcium bath of different concentration [Ca] (20, 40 and 80mM as indicated) show distinct, clearly-resolved, conduction bands M0, M1, and M2 for which there are respectively zero, one, or two calcium ions trapped saturately at the selectivity filter. (b) The peaks in conduction correspond to transitions of occupancy $P$ between these saturated levels.} \label{fig:ca_bands}
\end{figure}


\begin{figure}[t]
\includegraphics[width=1.0\linewidth]{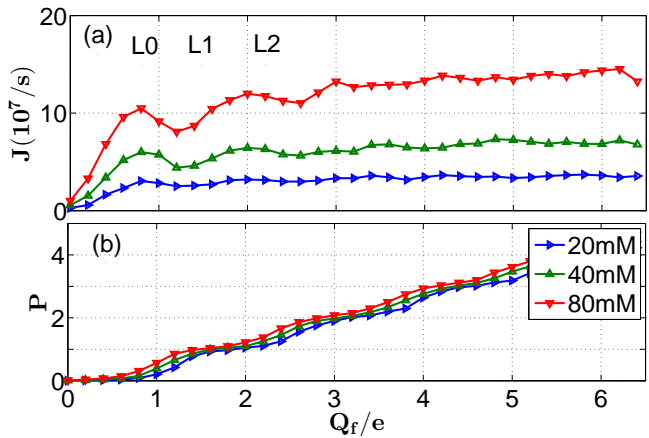}
\caption{(Color online)  BD simulations showing sodium conduction and occupancy bands of the generic ion channel model. (a) Plots of the sodium ionic current $J$ as a function of the fixed charge $Q_f$ at the selectivity filter for a pure sodium bath of different concentration [Na] (20, 40 and 80mM as indicated) show broadened conduction bands L0, L1, and L2 for which there are respectively zero, one, or two calcium ions trapped saturately at the selectivity filter.  (b) The broad conduction peaks still correspond to transitions of occupancy $P$ between these saturated levels.} \label{fig:na_bands}
\end{figure}


\begin{figure}[t]
\includegraphics[width=1.0\linewidth]{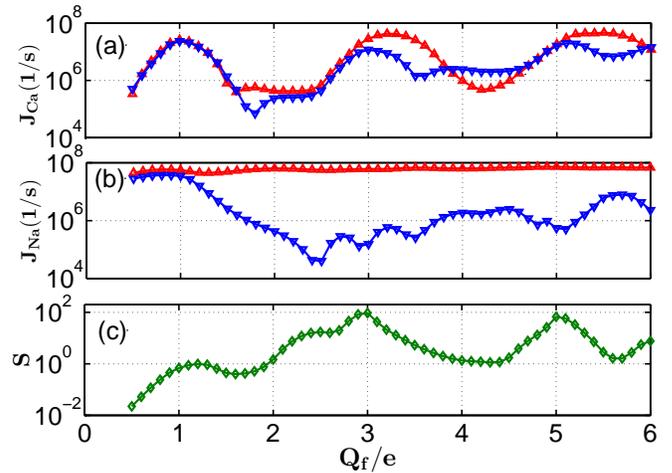}
\caption{(Color online)  BD simulations showing conduction and selectivity of the generic ion channel model in mixed and pure baths (partly reworked from \cite{Kaufman:13a}). Note the logarithmic ordinate scale. (a) The calcium current $J_{Ca}$ as a function of the fixed charge $Q_f$ at the selectivity filter in the mixed bath (blue point down triangles) shows significant attenuation for bands M1 and M2 as compared with the pure bath (red point-up triangles).  (b) The sodium current $J_{Na}$ as a function of $Q_f$ in the mixed bath (blue point-down triangles) exhibits progressive blockage by calcium as compared to the pure bath (red point-up triangles). (c) A plot of the selectivity ratio $S=J_{Ca}/J_{Na}$ for the mixed bath shows growth of selectivity for $Q_f$ above L0 and strong peaks corresponding to the M1 and M2 calcium bands.} \label{fig:bands}
\end{figure}

Fig.\ \ref{fig:ca_bands} and Fig.\ \ref{fig:na_bands} present results derived from Brownian dynamics simulations of permeation of the generic channel model by calcium and sodium ions in pure baths of different concentration.

Fig.\ \ref{fig:ca_bands}(a) shows the pronounced regular structure in the Ca$^{2+}$ ion current $J_{Ca}$ as a function of $Q_f$ for different Ca$^{2+}$ concentrations [Ca]. The structure consists of narrow regions of high conductance (conduction bands) $M0\approx1e$, $M1\approx3e$ and $M2\approx 3e$ separated by almost zero-conductance stop-bands. The peak separation $\Delta Q \approx 2e$ corresponds to the charge on one Ca$^{2+}$ ion. As shown in (b), the peaks in $J$  correspond to transition regions in the channel occupancy $P$, where $P$ jumps from one saturated integer value to the next one, while zero-conductance bands correspond to regions of constant integer $P$. The calcium conduction bands  correspond to the ion-exchange low-barrier phase transitions obtained analytically in \cite{Zhang:06}

Comparison of the $J$ and $P$ plots shows that conduction occurs at odd integer values of $Q_f/ze$, corresponding to non-zero total charge of the selectivity filter, whereas the non-conducting regions of constant $P$ correspond to even integer values of $Q_f/ze$, i.e.\ to the {\it neutralized state}. The neutralization approach will be further discussed in (Sec.\ \ref{Sec:neutral})

Fig.\ \ref{fig:na_bands} plots the equivalent results for (a) the sodium current and (b) the occupancy as functions of $Q_f$ in a pure NaCl bath with different concentrations. The current $J_{Na}$ exhibits weak local maxima that would appear to be analogous to the calcium conduction bands in Fig.\ \ref{fig:ca_bands}(a). We label them as L0, L1, L2, corresponding to the integer sodium occupancy $P_{Na}=0,1,2$ of the selectivity filter; these broad bands overlap and never fall to zero, making the sodium conductance relatively independent of $Q_f$. The separations of the L-band maxima are half the size of those in the calcium M-bands, reflecting the charge difference between Na$^+$  and Ca$^{2+}$ ions.

Values of maximal sodium and calcium currents are about 5--10 $\times 10^7$s$^{-1}$(10--20pA), corresponding roughly to the currents observed experimentally in sodium and calcium channels under physiological conditions \cite{Sather:03, Catteral:12}.

The appearance of the distinct conduction bands are attributable to ion-ion and ion-fixed charge electrostatic interaction and the discreteness of the multi-ion occupancy $P$ \cite{Yesylevskyy:05,Kharkyanen:10}. They are particularly well-defined for Ca$^{2+}$ in the calcium channel on account of the double-valence of Ca$^{2+}$, which enhances the electrostatic effects of valence selectivity \cite{Corry:05}. It will be shown explicitly below (see Secs.\ \ref{Sec:energetics1ion} and \ref{Sec:energetics2ions}) that both the calcium and sodium conduction bands correspond to resonance-like barrier-less conduction.

As already noted, conduction occurs at odd integer values of $Q_f$. They correspond to half-integer values of occupancy and non-zero total charge at the selectivity filter $Q_t=ze\dot P-Q_f$ , where $z$ is ion valence ($z=2$ for Ca$^{2+}$), whereas the non-conducting regions of constant $P$ correspond to the $Q_t\approx0$ {\it neutralised state}. The neutralisation approach is discussed  below (see Sec. \ref{Sec:neutral}).

The different positions of the conduction and forbidden bands for ions of different valence  (in pure baths) provide a basis for valence selectivity in a mixed bath. Fig.\ \ref{fig:bands} shows the transformations in the conduction bands that occur in a mixed bath ([Ca]=40mM, [Na]=30mM). It is evident in (a) that the calcium conduction and stop bands bands persist, although bands M1 and M2 are significantly attenuated in the mixed salt as compared to the pure one.

Fig.\ \ref{fig:bands}(b) makes clear that the sodium current exhibits a persistent block for $Q_f> {\rm M0} $ that intensifies progressively until the beginning of M1. This strong progressive blockade of the sodium current can be accounted for by the increase in the depth of the potential well (which is linear in $Q_f$) and the consequent exponential decrease of the escape rate of the blocking Ca$^{2+}$ ions.

Fig.\ \ref{fig:bands}(c) plots the selectivity ratio $S=J_{Ca}/J_{Na}$ showing how the channel is selective in favor of calcium in a mixed salt bath. The ratio starts from $S\approx 0.01$  for L0 (sodium-selective channel), increases to $S\approx 1$ (non-selective channel) for M0, drops again near L1 (sodium-selective channel) and then rises fast to the high selectivity peak of $S \approx 100$ for the calcium selective M1, corresponding to the L-type calcium channel. Note that the calcium selectivity peaks M1 and M2 are shifted to lower $Q_f$ related to the corresponding peaks in $J$ (cf.\ (a)). They  correspond to the thresholds of the transitions in $P$.

\subsection{Identification of selectivity bands in the calcium/sodium channels family}\label{Sec:identification}

We now try to relate the observed charge-ordered sequence of conduction and selectivity bands shown in Fig.\ \ref{fig:bands} to the behaviour exhibited by real channels in experiments. Calcium and sodium conduction and stop bands divide the $Q_f$ axis into a number of distinct regions differentiated by the type of Ca$^{2+}$/ Na$^+$ selectivity, i.e.\ by combination of 4 features: the Na$^+$ conductivity for a pure bath; the Ca$^{2+}$ conductivity for a pure bath; the existence and power of the divalent block; and the AMFE (i.e.\ calcium-selective current), for the mixed salt bath. Combining these features we can find several clearly differentiated $Q_f$ regions with distinct selectivity types related  to particular channels including wild-type, mutants, and artificial.

\begin{figure}[t]
\includegraphics[width=1.0\linewidth]{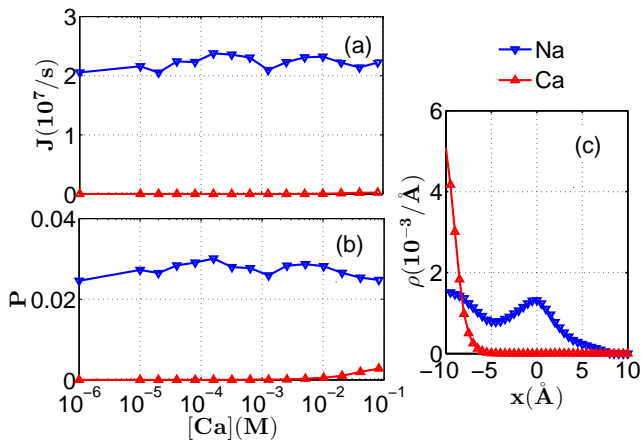}
\caption{(Color online) Band L0. BD simulations showing conduction and occupancy in a mixed salt bath with Na$^+$ (blue, point-down, triangles) and Ca$^{2+}$ (red, point-up, triangles); the lines are guides to the eye. (a) Sodium and calcium currents $J$ and (b) occupancies $P$ {\it vs.} the Ca$^{2+}$ concentration $[Ca]$ for $[Na]=30$\,mM. L0 shows moderate sodium conductivity without the divalent block corresponding to AMFE. (c) Mutual occupancy profiles for Na$^+$ (blue peaked curve) and Ca$^{2+}$ ions (red monotonic curve) show that the Ca$^{2+}$ ion cannot enter the channel.} \label{fig:conductionL0}
\end{figure}
Fig.\ \ref{fig:conductionL0}(a) shows that the band L0 ($Q_f \approx 0.5e$) demonstrates moderate sodium conductivity and near-zero calcium conductivity in a mixed bath (as in the pure baths), and that there is no divalent block or AMFE (plots (a), (b)). The near-zero $P_{Ca}$ shown in (b) is because calcium ions cannot overcome the self-potential barrier in order to enter the channel, as illustrated by the occupancy plot in (c). The results are almost independent of $[Ca]$. Thus L0 represents a non-blocking sodium-selective channel. We infer that this band is associated with the bacterial sodium NaChBac channel \cite{Yue:02}, which exhibits a similar type of selectivity; the same connection was also proposed recently by Corry \cite{Corry:13}. We can connect the L0 band with the DEKA inner ring of the mammalian Na$_{\rm v}$ sodium channel \cite{Vora:05, Catteral:12}.

\begin{figure}[t]
\includegraphics[width=1.0\linewidth]{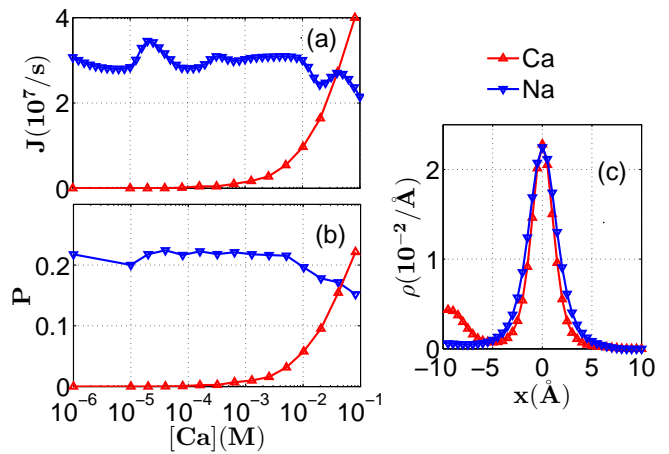}
\caption{(Color online) The non-selective band M0. BD simulations showing conduction and occupancy in a mixed salt bath with Na$^+$ (blue, point-down, triangles) and Ca$^{2+}$(red, point-up, triangles); the lines are guides to the eye. (a) Sodium  and calcium  currents $J$ and (b) occupancies $P$ {\it vs.} Ca$^{2+}$ concentration $[Ca]$ for $[Na]=30$mM.  M0 shows non-selective currents both in pure and mixed baths. (c) Mutual occupancy profiles for Na$^+$ and Ca$^{2+}$ ions show an absence of any blockade of Na$^+$ ions by the Ca$^{2+}$ ions, and a time-shared occupancy mode.} \label{fig:conductionM0}
\end{figure}

Fig.\ \ref{fig:conductionM0} shows that the M0 channel ($Q_f\approx 1e$) exhibits non-selective conduction and occupancy for both sodium and calcium (plots (a), (b)), and non-selective time-sharing mutual occupany profiles (plot (c)) and thus represents a non-selective cation channel. It may be identified with the non-selective cation channel described in \cite{Guinamard:12} or the OmpF channel \cite{Miedema:04}. The high calcium $J$ corresponds to barrier-less conductivity for Ca$^{2+}$ (see Sec. \ref{Sec:energetics2ions}).

\begin{figure}[t]
\includegraphics[width=1.0\linewidth]{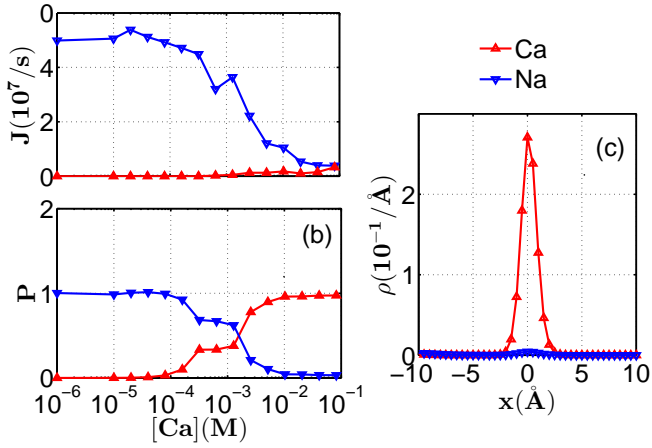}
\caption{(Color online) The sodium selective band L1. BD simulations showing conduction and occupancy in a mixed salt bath with  Na$^+$ (blue, point-down, triangles) and Ca$^{2+}$ (red, point-up, triangles); the lines are guides to the eye. (a) Sodium and calcium currents $J$ and (b) occupancies $P$ {\it vs.} Ca$^{2+}$ concentration $[Ca]$ for $[Na]=30$\,mM. L1 shows strong blockade without AMFE at $P_{Ca}=1$ with a threshold of $[Ca]_{50}\approx 1$\,mM. (c)  Mutual occupancy profiles for Na$^+$  and Ca$^{2+}$  ions show substitution and blockade of Na$^+$ ions by the first Ca$^{2+}$ ion which by itself completely occupies the channel.} \label{fig:conductionL1}
\end{figure}

\begin{figure}[t]
\includegraphics[width=1.0\linewidth]{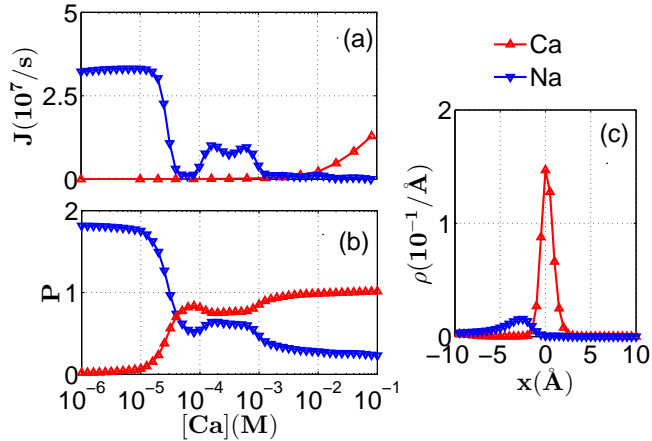}
\caption{(Color online) BD simulations showing AMFE in a mixed salt bath for the M1 calcium channel (reworked from \cite{Kaufman:13a}) with Na$^+$ (blue, point-down, triangles) and Ca$^{2+}$ (red, point-up, triangles); the lines are guides to the eye. (a) Sodium  and calcium currents $J$ and (b) occupancies $P$ {\it vs.} the Ca$^{2+}$ concentration $[Ca]$ for $[Na]=30$\,mM.  M1 shows strong blockade and AMFE at $P_{Ca}=1$, with a threshold of $[Ca]_{50}\approx\,30\mu$M. (c) Mutual occupancy profiles for Na$^+$ and Ca$^{2+}$  show blockade of Na$^+$ ions by one Ca$^{2+}$ ion.} \label{fig:AMFE_M1}
\end{figure}

Results for the double-occupied sodium band L1 ($Q_f\approx 1.5-2.0e$) are plotted in Fig.\ \ref{fig:conductionL1}. As shown in (a) and (b), this band exhibits high conductivity for pure sodium, zero conductivity for pure calcium, and blockade of the sodium current by calcium. Onset of the blockade occurs at $[Ca]_{50} \approx 1$mM after the first Ca$^{2+}$ ion has occupied the selectivity filter: $P_{Ca} \rightarrow 1$.  The mutual occupancy profiles for Na$^+$ and Ca$^{2+}$  shown in (c) demonstrate full substitution of Na$^+$ ions by the first Ca$^{2+}$ ion.  It is thus a sodium-selective channel that is subject to divalent block. This kind of selectivity corresponds to the wild type mammalian sodium channel \cite{Schlief:96,Hille:01} and it relates to an outer EEDD  ring of residues \cite{Vora:05, Catteral:12}, similar to the four-glutamate ring discovered in the bacterial NavAb channel.

The narrow calcium selectivity peaks M1 ($Q_f\approx 3e$) and M2 ($Q_f\approx 5e$) (Fig.\ \ref{fig:bands}(c)) exhibit highly non-selective conductivity in a pure bath, and strong divalent blockage of sodium, followed by AMFE. This kind of selectivity is a trade-mark of calcium channels \cite{Sather:03} identified with the wild-type L-type and RyR calcium channels, respectively \cite{Kaufman:13a}.

Fig.\ \ref{fig:AMFE_M1}  presents the dependences of $J$ and $P$ on $[Ca]$ for the M1 band in a mixed salt configuration. As shown in (a),(b), M1 shows a strong blockade of the current $J_{Na}$ of Na$^+$ ions with its onset at $[Ca]_{50} \approx 30\mu$M. The blockade occurs after the first Ca$^{2+}$ ion has occupied the selectivity filter: $P_{Ca} \rightarrow 1$ as shown in (b). The mutual occupancy profiles for Na$^+$ and Ca$^{2+}$ shown in (c) also indicate blockade of Na$^+$ ions by the first Ca$^{2+}$ ion. This is a calcium-selective channel with single-ion calcium block. Strong blockade with a relatively low onset agrees qualitatively with the observed properties of the L-type channel \cite{Sather:03}. The value of $Q_f$, and the conduction mechanism for M1, also correspond to the model \cite{Corry:01} of the L-type channel (EEEE locus).

\begin{figure}[t!]
\includegraphics[width=1.0\linewidth]{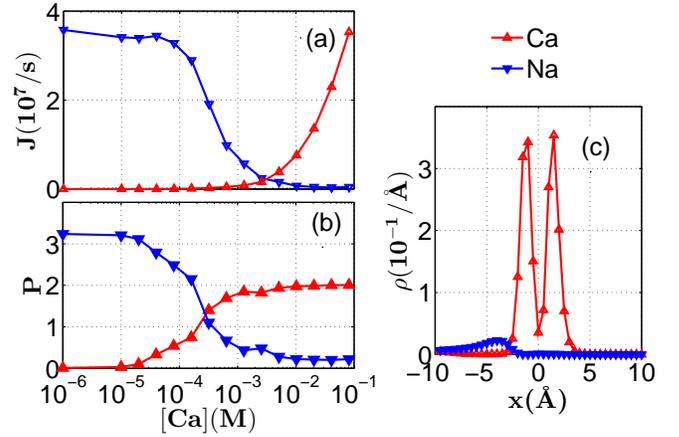}
\caption{((Color online) AMFE in a mixed salt bath for the M2 channel (reworked from \cite{Kaufman:13a}) with Na$^+$ (blue, point-down, triangles) and Ca$^{2+}$ (red, point-up, triangles); the lines are guides to the eye. (a) Sodium and calcium currents $J$ and (b) occupancies $P$ {\it vs.} the Ca$^{2+}$ concentration $[Ca]_{50}$ for $[Na]=30$\,mM. M2 shows strong blockade and AMFE at $P_{Ca}=1$ with a threshold of $[Ca]_{50}\approx\,150\mu$M. (c) Mutual occupancy profiles for Na$^+$  and Ca$^{2+}$ ions show blockade of Na$^+$ ions by a pair of Ca$^{2+}$ ions.} \label{fig:AMFE_M2}
\end{figure}

Fig.\ \ref{fig:AMFE_M2} provides similar information for the M2 band, again in a mixed salt configuration. As shown in (a),(b), there is a strong blockade of the current $J_{Na}$ of Na$^+$ ions with its onset at $[Ca]_{50} \approx 150\mu$M after two Ca$^{2+}$ ions have occupied the selectivity filter: $P_{Ca} \rightarrow 2$. This is a calcium-selective channel with double-ion calcium block. Divalent blockade with a relatively high onset and strong calcium current agrees qualitatively with the observed properties of the RyR calcium channel \cite{Gillespie:08} and with the TPRV6 channel \cite{Owsianik:06}.

We thus arrive at the full identification scheme presented in Table \ref{tab:identification}; it represents a completed version of the partial table in \cite{Kaufman:13a}.

Within the framework of our scheme, an increase of fixed negative charge at the selectivity filter leads to an increase of calcium selectivity with the strict sequence: L0 (sodium selective, non-blocking channel) $\rightarrow$ M0 (non-selective cation channel) $\rightarrow$ L1 (sodium selective channel with divalent block) $\rightarrow$ M1 (calcium selective channel with divalent, blockade, AMFE). And {\it vice versa}, a decrease in the negative charge should change the selectivity from an L-type calcium channel to sodium, and from sodium to non-selective. The sodium L1 blocking channel holds an intermediate position in the $Q_f$ spectrum between the non-selective M0 channel and the calcium-selective M1 channel. A similar increase of selectivity with increasing $Q_f$ was obtained in  \cite{Miedema:04,Csanyi:12} but without the sharp selectivity peak at M1.

Comparison between the L1 and M1 conductivity-selectivity behavior shows a close similarity of their blockade mechanisms, but a significant difference between their calcium conductivities. Unlike the calcium-selective band M1, which exhibits narrow selectivity peaks, the sodium-selective band shows conduction/selectivity properties over a relatively wide range of $Q_f$ around L1. This difference can be explained as being the result of barrier-less Ca$^{2+}$ conductivity appearing in the narrow M1 band (see Sec.\ \ref{Sec:energetics1ion}). In some sense, a sodium channel can be described as a sub-optimal calcium channel: the Ca$^{2+}$ ion blocks the Na$^{+}$ current, but the Ca$^{2+}$-Ca$^{2+}$ knock-on mechanism does not work.

\begin{table*}[t]
\caption{Identification of conduction and selectivity bands of the model with known ion channels.}\label{tab:identification}

\begin{tabular}{|p{1.8cm}|p{1.8cm}|p{1.5cm}|p{1.5cm}|p{1.5cm}|p{1.5cm}|p{3.5cm}|p{2.3cm}|}\hline
               &          & \multicolumn{2}{c}{\bf Pure salt bath} \vline & \multicolumn{2}{c}{\bf Mixed salt bath} \vline & & \\\cline{3-4} \cline{5-6}
\vspace{-0.5cm}{\bf Conduct- ion bands} & \vspace{-0.5cm}$\approx${\bf Fixed charge {\itshape (e)}} &   &   &  &  & \vspace{-0.5cm}{\bf ~~~~~~~~Channels} & \vspace{-0.5cm}{\bf Locus / nominal charge}\\
& & \vspace{-0.7cm}Na$^+$ ~~~~~~ current & \vspace{-0.7cm}Ca$^{2+}$ ~~~~~~ current & \vspace{-0.7cm}Blockade & \vspace{-0.7cm} AMFE & & \\ \hline
L0 & 0.5 & Moderate & Low      & No & No &  Nav sodium \cite{Hille:01} & DEKA($1e$)\\
M0 & 1        & Moderate & Moderate & No & No & Non-selective
                                                            OmpF \cite{Miedema:04} & RRRDE(2e)\\
L1 &     1.5-2.0  & Low      & High     & Yes & No & NaChBac \cite{Yue:02}, NavAB\cite {Payandeh:11} & EEEE(4e)\\
M1             & 3        & High     & High     & Yes & Yes & L-type calcium  \cite{Sather:03}  & EEEE($4e$)\\
M2             & 5        & High     & High     & Yes & Yes & RyR calcium \cite{Gillespie:08} & DDDD(ED)($6e$)\\
\hline
\end{tabular}

\end{table*}

\begin{table*}[t]
\caption{Mutation-induced selectivity transitions in calcium and sodium ion channels.}\label{tab:transitions}

\begin{tabular}{|p{6.0cm}|p{5.5cm}|p{3cm}|}\hline
{\bf Channel transformation}        & {\bf Locus changes/nominal charges}   & {\bf Band transition/charges}\\ \hline
Na$_{\rm v}$ sodium $\rightarrow$ calcium selective \cite{Heinemann:92}
                                    & DEKA ($1e$) $\rightarrow$ DEEA($3e$)
                                                                     & L0($0.5e$) $ \rightarrow$ M1($3e$)\\
Ca$_{\rm v}$ calcium $\rightarrow$ sodium selective mutant \cite{Yang:93}
                                    & EEEE ($4e$) $\rightarrow$ DEDA ($3e$)
                                                                     & M1 ($3e$)$ \rightarrow$ L1 ($1.5e$)\\

Ca$_{\rm v}$ calcium $\rightarrow$ nonselective mutants $\rightarrow$ sodium-selective mutant \cite{Tang:93}
                                    & EEEE $\rightarrow$ EEEA(Q) $\rightarrow$ EEEK  $\rightarrow$ EEKA 
                                                                     & M1 $\rightarrow$ L1 $\rightarrow$ M0$\rightarrow$ L0\\

Na$_{\rm v}$ sodium $\rightarrow$ numerous mutants with different loci \cite{Schlief:96}
                                    & DEKA  $\rightarrow$ DEKE $\rightarrow$ DEEA $\rightarrow$ EEEE $\rightarrow$ DEEE
                                                                     & L0$\rightarrow$ L1$ \rightarrow$ M1$\rightarrow$ M1\\
Nonselective OmpF porin $\rightarrow$ calcium selective mutant \cite{Miedema:04}
                                    & RRRDE ($2e$)$\rightarrow$ DEEE ($4e$)$\rightarrow$ (L)AEA ($7e$)
                                                                      & M0($1e$)$\rightarrow$ M1($3e$)$\rightarrow$ M2($5e$)\\
\hline
\end{tabular}

\end{table*}

\subsection{Mutation-induced transitions between selectivity bands}\label{Sec:mutations}

The identification of the selectivity bands in the model with real channels (wild type and mutants) allows us to establish a model ``charge scale" for different channels and different mutations (Table \ref{tab:transitions}), and to compare it with the more conventional charge scale based on the nominal charges of amino acid side chains at normal electrolyte pH values within a channel. Such a scale allows us to describe/predict the known/possible results of mutations leading to substitutions of residues at the selectivity filter with residues of different charge, or to the elimination of particular residues. The scale based on our model is generally similar to that suggested in \cite{Miedema:04,Csanyi:12}, albeit with different charges for some residues and channels.

Typically, the effective charge $Q_f$ in our model appears  to be less than sum of the nominal charges of the residues assigned to the locus in question. Our simulations give M1=3$e$ for conservative EEEE locus of L-type calcium channel (nominal charge $4e$), and $L0=0.5e$ for the DEKA locus of the Na$_{\rm v}$ sodium channel (nominal charge $1e$), in agreement with earlier BD simulation results for calcium and sodium channel \cite{Corry:01, Vora:05}.

These differences could be partially (and speculatively) related to possible difference in ionisation states of residues \cite{Nonner:98, Varma:04}. Multiple amino acids  with carboxylate groups have to be placed in close proximity to establish the rings of high negative charge in the selectivity filters considered in the model; protonation at neutral pH would either reduce the effective charge on the residues \cite{Nonner:98, Varma:04} or leave them unchanged. Mutant studies with residues eliminated confirm that there is significant protonation of the ring of glutamates \cite{Sather:03}.

Our computation of the effective fixed charge might also be misleading in some cases, because the model includes only the fixed charge of the filter locus (the ring). For instance, the assignation of effective charges to the RyR DDDD locus takes no account of the fact that the pore of the RyR channel is lined by a total of 20 negatively-charged residues, not only by the D4899 charges of the 4 alpha-subunits \cite{Gillespie:08}.

Thus the most obvious reason for differences between the conventional/accepted charges of residues and the band-derived values of $Q_f$ is the simplicity (and generality) of our model, and possible differences between the model parameters and the real channel structure parameters (which are sometimes unknown). Parametric studies of the model have shown reasonable robustness in the positions of the conduction bands to variations of $R$ and $L$ in the selectivity filter, to the width $H$ of the charge ring and to the membrane potential $V$ \cite{Kaufman:13a}. However the model's effective scale of $Q_f$ could be affected by e.g.\ the assumed hydration model and the assumed radius of the charged ring. The bands shift upwards in $Q_f$ with hydration barrier growth and downwards with increasing radius of the charged ring.  In this work, we do not use fitting procedures, and nor do we assume modified values for the charges on the residues.
We just take the $Q_f$ band values that emerge as effective values related but not equal to the real charges.

\begin{figure}[t!]
\includegraphics[width=1.0\linewidth]{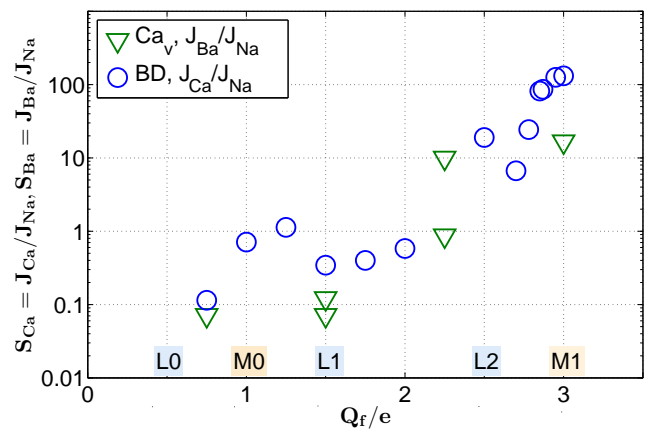}
\caption{(Color online) Mutation-induced increase of divalent (calcium or barium) over sodium selectivity $S_{Ca}=J_{Ca}/J_{Na}, S_{Ba}=J_{Ba}/J_{Na}$, comparison of BD simulations with experiment. The generic model predicts fast growth of $S_{Ca}$ from 0.01 for L0 to 100 for M1 (blue circles). The simulation results are in reasonable agreement with the experimental results for $S_{Ba}$ in
Ca$_{\rm V}$ to sodium selective mutants (green triangles, where $Q_f$ is scaled by M1.) \cite{Tang:93}} \label{fig:NavChBac}
\end{figure}

Fig. \ref{fig:NavChBac} compares the BD simulation results for Ca/Na selectivity $S_{Ca}=J_{Ca}/J_{Na}$    with experimental data for Ba/Na selectivity $S_{Ba}=J_{Ba}/J_{Na}$.  Our generic model predicts fast growth of $S$ from 0.01 for L0 to 100 for M1 (blue circles). 
The green open triangles  are from mutation studies of Ca$_{\rm v}$ channel and its less-charged mutants (EEEE ($4e$) $\rightarrow$ EEEA(Q) ($3e$)$\rightarrow$ EEEK ($2e$) $\rightarrow$ EEKA ($1e$)). They demonstrate a clear dependence of $S_{Ba}=J_{Ca}/J_{Na}$ on $Q_f$ \cite{Tang:93} in agreement with the predictions of our model.  Consequently we identified these channels with bands from M1 down to L0.
Table \ref{tab:transitions} lists the above-mentioned and some other known mutation transformations together with their attributions within the framework of our model.

The recently investigated NALCN channel is a member of the family of ion channels with four homologous repeat domains that include voltage-gated calcium and sodium channels. NALCN appears in two variants with selectivity filter residues that resemble either calcium channels (EEEE) or sodium channels (EKEE or EEKE), controlled by a single gene \cite{Senatore:13}. We can tentatively identify the EEKE channel with the L1 band and EEEE with the M1 band. Reversible transformations between these states can be identified as L1$\leftrightarrow$M1 transitions.

An appropriate point mutation of the DEKA sodium channel ($Q_f \approx 1e$) converts it into a calcium-selective channel with a DEEA locus \cite{Heinemann:92}. Our scheme identified this result with the L0 ($Q_f=0.5e$)$\rightarrow$M1 ($Q_f=3e$) transition.

The essentially non-selective bacterial OmpF porin with its RRRDE locus can be converted into a Ca$^{2+}$-selective channel by the introduction of two additional glutamates in the constriction zone; the resultant mutant contains a DEEE-locus and exhibits an Na$^+$ current with a strongly increased sensitivity to 1\,mM Ca$^{2+}$. Another OmpF mutant with formal net charge $Q_f$=7e demonstrating weaker AMFE and smaller selectivity to Ca$^{2+}$  was identified as an analogue of the RyR channel  \cite{Miedema:04}. We can identify this transformations with the M0 ($Q_f=1.0e$)$\rightarrow$M1 ($Q_f=3e$)$\rightarrow$M2 ($Q_f=5e$) transitions.

Thus our identification scheme provides straightforward explanations for the outcomes of several mutant studies. Some results still seem to lie outside the scope of our model, however, e.g.\ the change of the ions' permeation/selectivity properties by a simple permutation of the residues at the selectivity filter\cite{Schlief:96}.

The calcium-selective M1 band exhibits a narrow resonance-like selectivity peak. We can conclude that  any single mutation of the calcium channel which influences $Q_f$ should destroy its specific calcium selectivity.  It corresponds well  with the facts that the EEEE signature for the L-type channel is highly conserved \cite{Sather:03} and that mutations in the genes responsible for this selectivity filter motif lead to numerous diseases \cite{Burgess:99}.

The resonance-like nature of calcium selectivity is particularly interesting in connection with the recently discovered NavAb sodium channel which possesses the same EEEE locus as the calcium L-type channel but exhibits sodium-selective permeation behavior \cite{Payandeh:11,Payandeh:12}. In the context of our model, this paradox could be explained in terms of a geometry difference (relatively small length of selectivity filter, or large radius), here we should bear in mind that the bands disappear when $L$ decreases to 8\,\AA ~ or $R$ reaches to 4.5\,\AA ~ \cite{Kaufman:13a}, so that calcium selectivity could drastically decrease.
Another plausible explanation relates to possible variations in the protonation of residues for different channels and therefore to slightly different effective charge for nominally the same loci \cite{Cymes:05, Varma:04, Finnerty:12}. Due to the narrowness of the calcium-selective M1 band even small changes of total charge could convert it to sodium channel. Clarification of these questions will require further experimental research and more detailed simulations.

\subsection{Energetics of single-ion conduction and selectivity bands L0 and M0}\label{Sec:energetics1ion}

We now  investigate the energetics of calcium and sodium conductivity in our model,  and show explicitly  that a barrier-less permeation mechanism underlies the appearance of the conduction and selectivity bands.

\begin{figure}[t]
\includegraphics[width=1.0\linewidth]{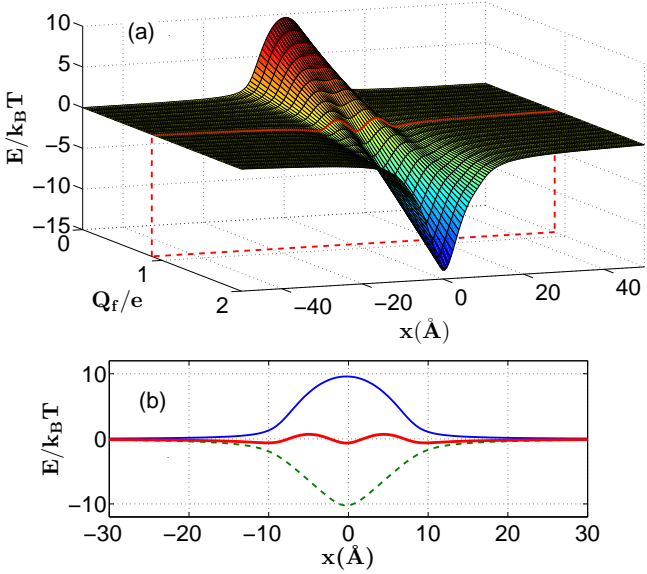}
\caption{(Color online) The appearance of a barrier-less path for the channel M0. (a) The electrostatic potential energy profile along the channel's $x-$axis is plotted {\it vs.} the fixed charge $Q_f$. The energy differences across the profile are minimal at a particular value of $Q_f=Q_{opt}$. (b) This optimal profile for permeation (red) appears as the result of a balance between repulsion by the dielectric boundary force (blue) and attraction to the fixed charge (green, dashed).} \label{fig:lowbarrierM0}
\end{figure}

Fig.\ \ref{fig:lowbarrierM0} shows that barrier-less conductivity for cations of particular valence (the calcium M0 band is drawn) appears as the result of a balance between the self-repulsion of the dielectric boundary force and the electrostatically amplified attraction to the negative fixed charge. The self-repulsion is proportional to $Q_i^2$, whereas the attraction is proportional to $Q_i\times Q_f$. Thus a variation of $Q_f$ can significantly change the resultant profile. This kind of barrier-less selectivity was suggested earlier for the Gramicidin channel \cite{Levitt:78,Nadler:03,Roux:04}.

Fig.\ \ref{fig:lowbarrierM0}(a) illustrates the fact that, for small $Q_f<$\,M0, self-repulsion dominates and the channel is not permeable by any ion; and that, for large $Q_f$ ($Q_f>\,$M0), attraction dominates and the ionic dynamics is then controlled by Kramers escape from a deep potential well, with an exponential dependence on its depth, $\Delta E$. Between these two extremes an optimal point $Q_{opt}$ ($Q_{opt}=0.9e$ for Ca$^{2+}$ ions) exists where $\Delta E=|E_{max}-E_{min}|$ is minimized with the appearance of an almost barrier-less ($\Delta E \sim k_BT$) profile for the moving ion.  Sodium ions exhibit a similar pattern but with $Q_{opt}=0.45e$ providing for valence selectivity between monovalent Na$^{+}$ and divalent Ca$^{2+}$  ions (see also below, Fig.\ \ref{fig:electroBrownM0}).

Fig.\ \ref{fig:lowbarrierM0}(b) shows  that  for $Q_f=M0$ the self-potential barrier of the dielectric boundary force is balanced by electrostatic attraction to the fixed charge $Q_f$, resulting in a low barrier with $\Delta E \sim  k_BT$.

To compare the results of electrostatic calculations and BD simulations, we introduce a simplified kinetic model that allows us to connect the energy difference along the energy profile with the current $J$ and occupancy $P$.

For a singly-occupied channel, and assuming that there is no back-flow, we get a linear dependence of $J$ on $P$ in the Kramers rate approximation : $J=k_0 \times P$ where $k_0$ stands for the escape rate. Coulomb interactions between the ion inside the channel, and ions in the bath and at the mouth, cause $k_0$ to be dependent on concentration \cite{Nadler:04, Tindjong:12a} and so lead to deviations of $J$ from a linear dependence on $P$.

We assume the generalized Kramers equation for $k_0$ in the vicinity of M0 or L0: $k_0 \approx D/L^2\exp(-\Delta E/k_BT)$ and get the resultant expression for the current $J$:

\begin{figure}[t]
\includegraphics[width=1.0\linewidth]{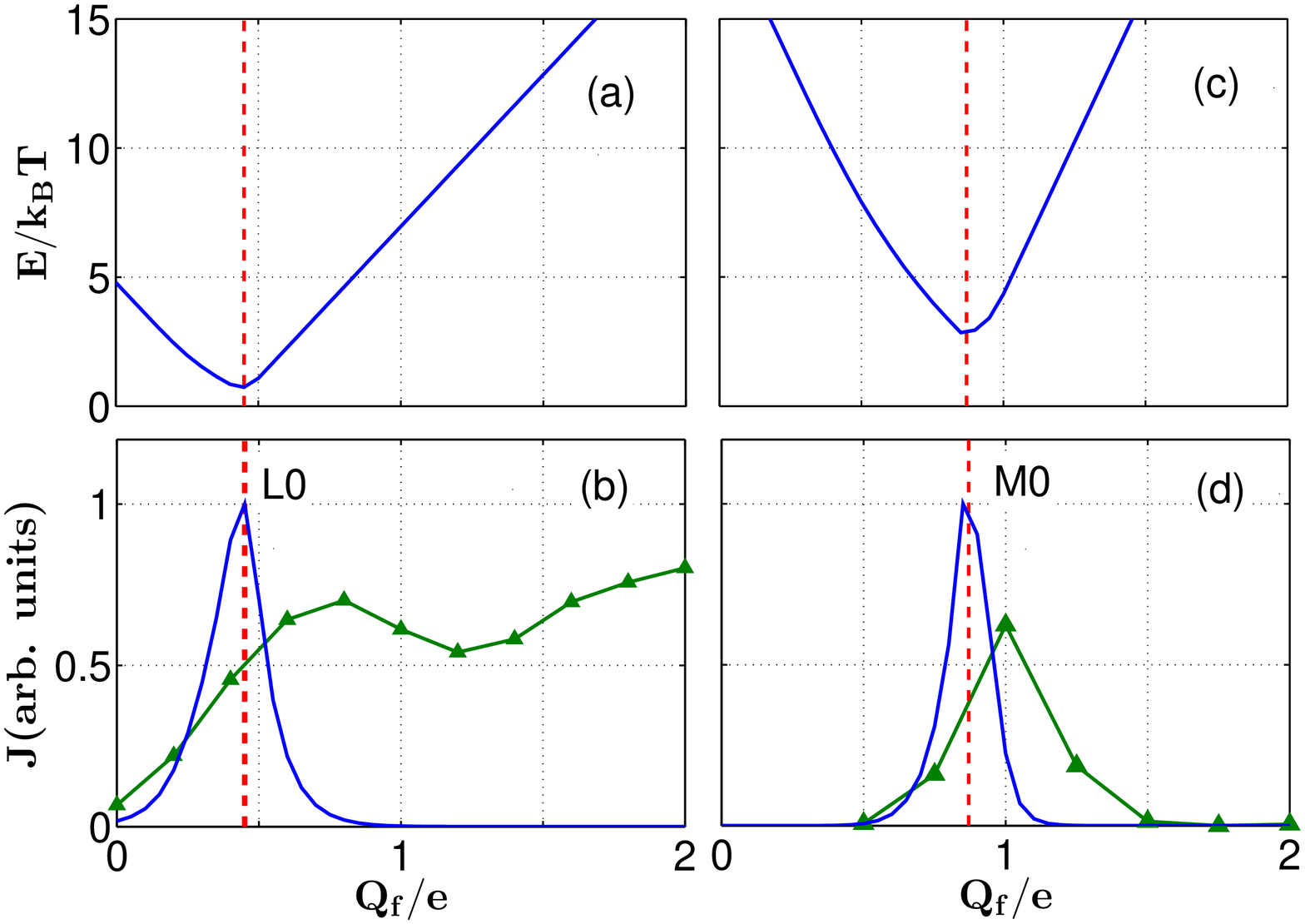}
\caption{(Color online) Energetics and Brownian dynamics of the single-ion permeation in the sodium L0 and calcium M0 bands. (a) The L0 band potential energy {\it vs.} the fixed charge $Q_f$. The energy difference along the profile shows a clear minimum at $Q_{opt}=0.45e$. (b) The peak in the sodium current $J$ {\it vs.} $Q_f$ calculated from electrostatics (blue curve) lies relatively close to the BD-simulated L0 peak (green point-up triangles, detrended). (c) The M0 potential energy {\it vs.} the fixed charge $Q_f$. The energy difference along the profile show a sharp minimum at $Q_{opt}=0.87e$. (d) The peak in the calcium current $J$ {\it vs.} $Q_f$ calculated from electrostatics (blue curve) lies close to the BD-simulated M0 peak. (green point-up triangles).} \label{fig:electroBrownM0}
\end{figure}
\begin{figure}[t]
\includegraphics[width=1.0\linewidth]{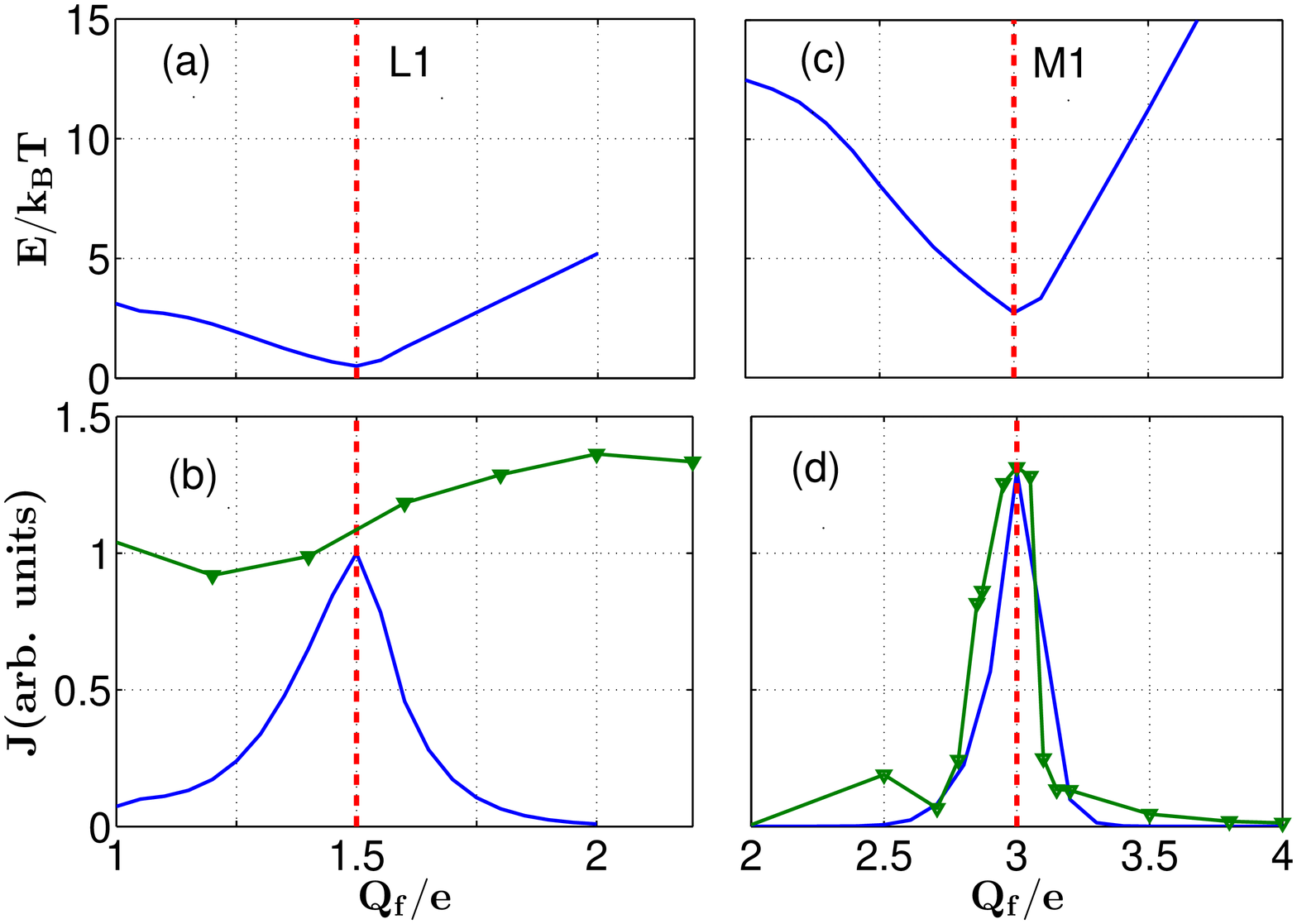}
\caption{(Color online)  Energetics and Brownian dynamics for double-ion permeaion of the sodium L1 and calcium M1 bands. (a) The L1 band potential energy {\it vs.} the fixed charge $Q_f$. The energy difference along the profile shows a wide minimum at $Q_{opt}=1.5e$. (b) The peak in the sodium current $J$ {\it vs.} $Q_f$ calculated from electrostatics is shifted down compared to the very weak BD-simulated conductance peak L1 (green point-down triangles). (c) The M1 calcium band potential energy {\it vs.} fixed charge $Q_f$. The energy difference along the profile show a deep minimum at $Q_{opt}=3e$. (d) The peak in the calcium current $J$ {\it vs.} $Q_f$ calculated from electrostatics (blue curve) lies close to the BD-simulated M1 peak in selectivity (green point-down triangles).}  \label{fig:electroBrownM1}
\end{figure}
\begin{equation}
J= k_0 P \approx D/L^2 \exp(-\Delta E/k_BT) \times P
\label{eq:Kramers}
\end{equation}

\noindent or, assuming that $P$=const,
\begin{equation}
J= J_0 \times \exp(-\Delta E/k_BT)
\label{eq:KramersM0}
\end{equation}
where $J_0$ is a reference current. We will use  (\ref{eq:KramersM0}) to compare the $J(Q_f)$ dependences obtained from electrostatics with the results of BD simulations.

\begin{figure}[t]
\includegraphics[width=1.0\linewidth]{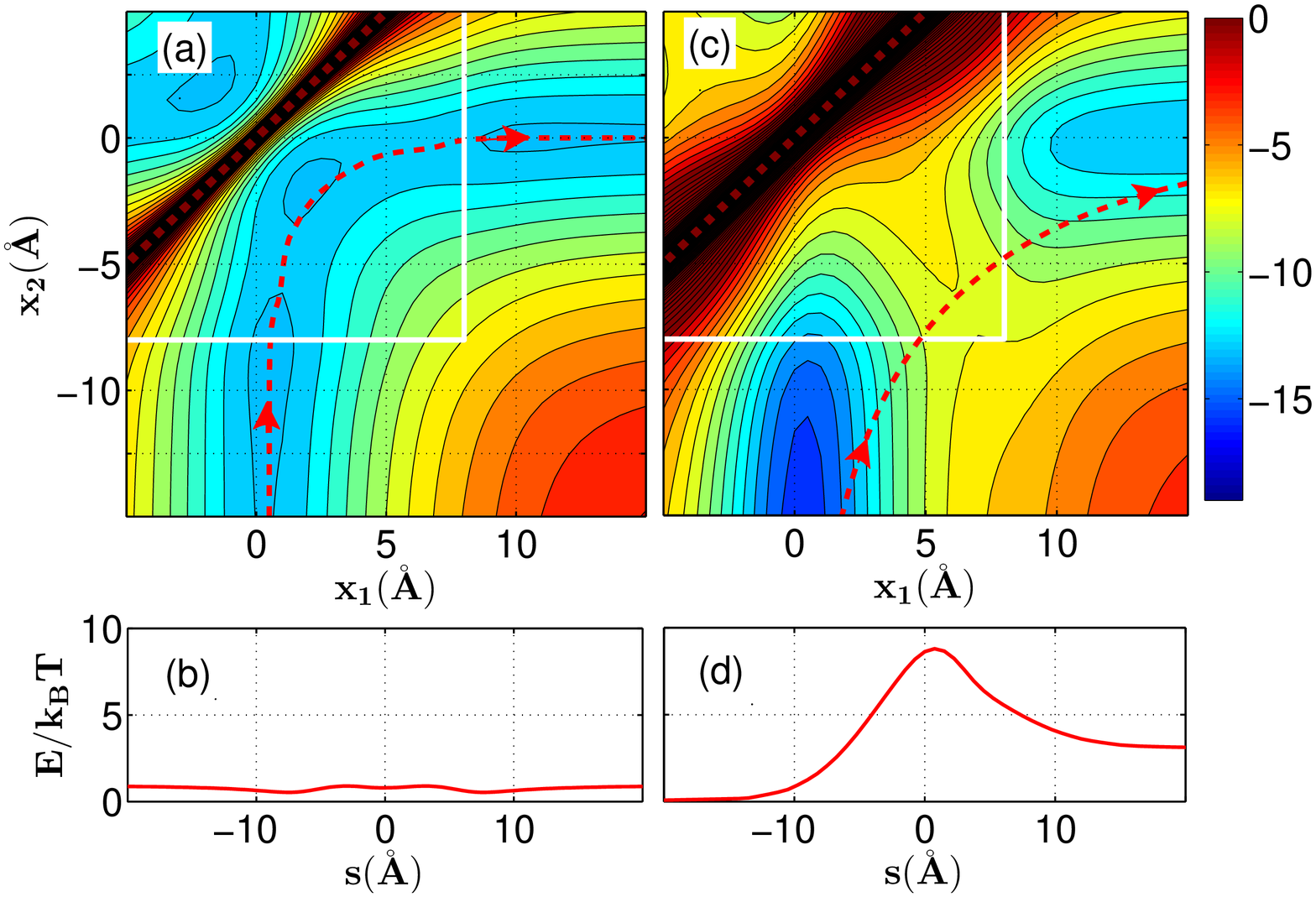}
\caption{(Color online) (a) Double-ion Na$^+$-Na$^+$ potential energy surface (PES) for the model channel with $Q_f=$L1 $(1.5e)$ , shown as a contour plot. The contour separation is 1$k_BT$, and the colorbar labels are in units of $k_BT$. The diagonal ridge in the upper-left corner represents the electrostatic barrier along the main diagonal of the map  $x_1=x_2$. The map area related to the selectivity filter is limited by the white lines. The optimal trajectory S (red dashed line) traverses two orthogonal valleys in the direction shown by the arrows and represents a ``knock-on'' event. The first ion initially captured at the center of the selectivity filter is pushed and substituted for by the second ion arriving at the channel mouth. (b) The potential energy $E$ profile along S represents almost barrier-less permeation. Plots (c),(d) show the same quantities for the heterogeneous Ca$^{2+}$-Na$^+$ double-ion: the binding site is initially occupied by a Ca$^{2+}$ ion that should be pushed by a Na$^+$ ion. The optimal trajectory S navigates via two valleys separated by a saddle which creates an intermediate potential barrier ($\Delta E \approx 8 k_BT$), corresponding to divalent blockade of Na$^+$.}
\label{fig:doubleionPESmapL1}
\end{figure}

\begin{figure}[t]
\includegraphics[width=1.0\linewidth]{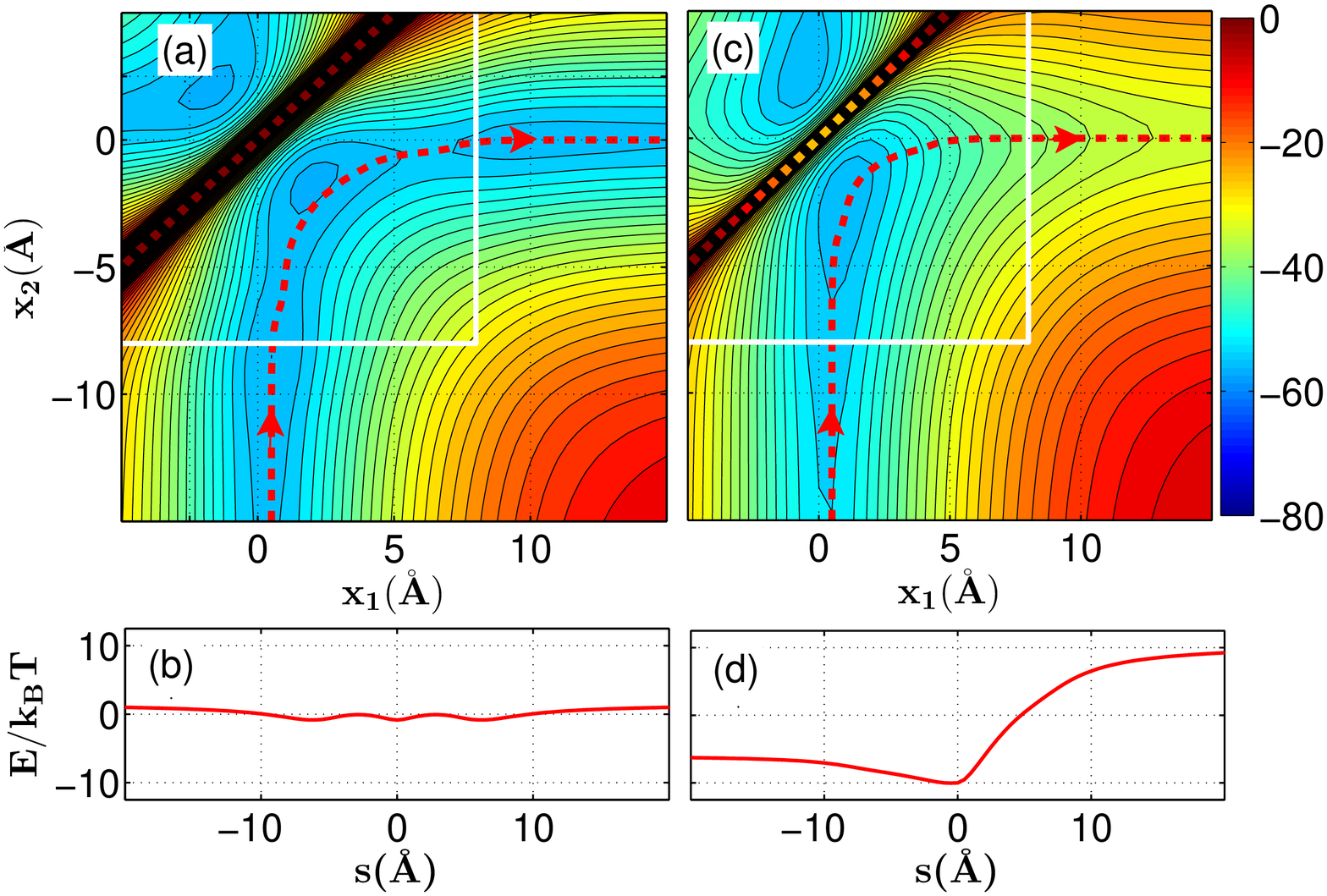}
\caption{(Color online)
(a) Double-ion Ca$^{2+}$-Ca$^{2+}$ potential energy surface (PES) for the model channel with $Q_f=$ M1 $(3.0e)$, shown as a contour plot. The contour separation is 2$k_BT$, and the colorbar labels are in units of $k_BT$. The diagonal ridge in the upper left corner represents the electrostatic barrier along the main diagonal of the map  $x_1=x_2$. The map area related to the selectivity filter is limited by the white lines. The optimal trajectory S (red dashed line ) traverses two deep ($\approx 60k_BT$) orthogonal valleys in the direction shown by the arrows, and represents a ``knock-on'' event. The first ion initially captured at the center of the selectivity filter is pushed out and substituted for by the second ion arriving at the channel mouth. (b) The potential energy $E$ profile along S demonstrates almost barrier-less permeation ($\Delta E<2kBT$). Plots (c),(d) show the same quantities for the heterogeneous Ca$^{2+}$-Na$^+$ double-ion: the binding site is initially occupied by a Ca$^{2+}$ ion that should be pushed by a Na$^+$ ion. The optimal trajectory S navigates via a deep valley ended by high potential barrier ($\Delta E \approx 20k_BT$), corresponding to deep divalent blockade of the Na$^+$ current. }
\label{fig:doubleionPESmapM1}
\end{figure}

Fig.\ \ref{fig:electroBrownM0} compares the energetics and BD results for the singly-occupied sodium L0 and calcium M0 bands.  The electrostatically calculated dependences of $\Delta E$ on $Q_f$ reveal sharp minima (a) at $Q_{opt}=0.45e$ for the L0 and (c) at $Q_{opt}=0.9e$ for the M0. The calcium band M0 exhibits a sharper $Q_f$ dependence because of the twice-larger charge on Ca$^{2+}$.

Figs.\ \ref{fig:electroBrownM0}(b),(d) compare the shapes and positions of the electrostatically calculated conductivity peaks ($J\sim J_0\exp( \Delta E/k_BT))$ with those obtained from the BD simulations. Their positions are in reasonable agreement, although the BD simulated peaks are shifted towards higher $Q_{f}$, probably due to kinetic effects related to (\ref {eq:Kramers}) and to the changing values of $P$ and $k_0$. These results support our inference that the resonance-like L0 and M0 bands maxima are attributable to barrier-less conduction. The conduction maximum for L0 shown in (b) is broadened towards higher $Q_f$ due partly to its overlap with L1 and partly to the slower exponent in $J$ for sodium. It leads to non-selective conduction for M0, as discussed above.

\subsection{Barrier-less  double-ion trajectories for conduction bands L1 and M1}\label{Sec:energetics2ions}
The double-ion sodium  selectivity band L1 is identified with the outer ring of the Na$_{\rm v}$ sodium channel and with the NLCN sodium channel. The double-ion calcium band M1 is identified with the L-type calcium channel \cite{Corry:01,Kaufman:13a}.

Multi-ion conductance appears when the selectivity filter potential well becomes too deep (about 60$k_BT$ for Ca$^{2+}$ in the vicinity of M1) making the channel impermeable when occupied by just one ion. Instead, conduction events occur via a double-ion knock-on conduction mechanism. This mechanism is caused by the electrostatic interaction between simultaneously captured ions, a process that is particularly effective for divalent  Ca$^{2+}$ ions \cite{Hess:84,Corry:01}.

The interacting calcium or sodium ions move simultaneously, in a coordinated manner, enabling escape to occur over a potential barriers of minimal height on the 2D potential energy surface (PES) \cite{Ganesh:00,Gordon:12}. The PES approach allows one to describe double-ion conduction as the potential motion of a quasi-particle along an optimal stochastic trajectory on the PES \cite{Dykman:92a,Elber:95,Kharkyanen:10}, thereby reducing the problem of double-ion conduction to the case already discussed, i.e.\ the 1D movement of a particle (in this case a ``super-ion'') in an electrostatic field.

We exploit this approach to show explicitly that the resonance-like conduction and selectivity of the M1 calcium channel and the L1 sodium channel occur through a barrier-less, multi-ion, conduction mechanism. To study valence selectivity we construct both homogeneous Ca$^{2+}$-Ca, Na$^+$-Na, and heterogeneous Na$^+$-Ca$^{2+}$ double-ion PESs, find the optimal (minimal energy change) stochastic paths, and calculate the energy profiles along these paths (see below).

Fig.\ \ref{fig:electroBrownM1}(a) shows the calculated dependence of $\Delta E$ on $Q_f$ for the L1 band (DEKA sodium channel), revealing  a smooth  minimum at $Q_{opt}=1.5e$. These data are obtained from an analysis of optimal trajectories for the electrostatic PES. A comparison of the current calculated from Kramers' approximation with that obtained from the BD simulations is shown in (b). The BD simulated maximum in the sodium current at $Q_{opt}=2e$ is very weak and shifted up relative to the point of barrier-less conductance as shown in (b). The discrepancy can be attributed to kinetics effects and to the obviously strong overlap between the different sodium bands (see Fig.\ \ref{fig:bands}).

Fig.\ \ref{fig:electroBrownM1}(c) shows similar comparisons for the M1 band (L-type calcium channel): the calculated dependence of $\Delta E$ on $Q_f$ undergoes  a sharp minimum at $Q_{opt}=3e$. The current calculated from the Kramers approximation is compared with that obtained from the BD simulations in (d). The good agreement between the peaks confirms that the maximum of selectivity in the double-occupied M1 band corresponds to the point of barrier-less conductivity.


Fig.\ \ref{fig:doubleionPESmapL1} presents Na$^+$-Na$^+$ and Ca$^{2+}$-Na$^+$ PES maps, optimal trajectories, and corresponding energy profiles, for $Q_f$=L1, at the point of barrier-less conductivity. Plots (a),(b) show the Na$^+$-Na$^+$ PES map and energy optimal trajectory S corresponding to a knock-on event, navigating two orthogonal valleys from South to East on the PES. The energy profile along S is almost flat (the energy difference along the optimal path does not exceed 1$k_BT$) corresponding to fast, barrier-less, permeation.

In contrast, the optimal path on the heterogeneous Ca$^{2+}$-Na$^+$ PES for L1 (Fig.\ \ref{fig:doubleionPESmapM1}(c),(d)) passs via a saddle (where it is not well-defined) where it has to overcome a relatively high potential barrier by thermal activation. The latter is $\Delta E \approx 8 k_BT$ for a sodium ion trying to knock-on a calcium ion and $\Delta E \approx 6 k_BT$ for the opposite combination. This is the PES-language explanation for calcium blockade in the outer ring of the Na$_{\rm v}$ sodium channel.

The pattern for Ca$^{2+}$-Ca$^{2+}$ permeation is rather similar but all effects are much more pronounced.  Fig.\ \ref{fig:doubleionPESmapM1} presents Ca$^{2+}$-Ca$^{2+}$ and Ca$^{2+}$-Na$^+$ PES maps, optimal trajectories and the corresponding energy profiles for $Q_f$=M1, at the point of maximum Ca$^{2+}$/Na$^+$ selectivity. Fig.\ \ref{fig:doubleionPESmapM1} (a),(b) shows the Ca$^{2+}$-Ca$^{2+}$ PES map and the energy-optimal trajectory S, navigating two deep orthogonal valleys. The energy profile along S is again almost flat: the energy difference along the optimal path does not exceed 1-2 $k_BT$, corresponding to fast barrier-less permeation.

In contrast, the heterogeneous Ca$^{2+}$-Na$^+$ PES for M1 (Fig.\ \ref{fig:doubleionPESmapM1}(c),(d)) encounters an impermeable high potential barrier $\Delta E \approx 20 k_BT$ for a sodium ion trying to knock-on calcium ion., which would need to be overcome by thermal activation. This is the PES-language explanation for calcium blockade and AMFE in the EEEE calcium channel; the barrier for the opposite combination is considerably smaller, $\Delta E \approx 3 k_BT$, and it can be overcome by thermal activation i.e.\ a calcium ion can knock-on a sodium one.

It was shown rigiriusly in \cite{Kaufman:13c} that  calcium and sodium condiction and selectivity band correspond to barrier-less, double-ion, conduction for Ca$^{2+}$ ions and a deep blockade of  Na$^{+}$ ions, thereby resolving the selectivity {\it vs.} conductivity paradox.

In terms of our simple model, there is of course no essential difference between a biological ion channel and an artificial nanopore of similar geometry (radius $R$ and length $L$) and surface charge $Q_f$. Such nanopores may be expected to demonstrate similar conductivity and selectivity features and a number of practical applications can be envisaged.

\section {Charge neutralisation and valence selectivity}\label {Sec:neutral}

The pattern of conduction and occupancy bands revealed by BD simulations and confirmed by electrostatics appears as a set of equidistant (periodic) $J$ peaks coinciding with steps in $P$ with a period related to the ionic charge $ze$ and shifted from zero by a half-period $Q_f=ze/2$. We now offer a simplified (and non-rigorous) explanation of this phenomenon based on the idea of sequental neutralisation of the fixed negative charge $Q_f$ by the capture of positive ions.

The zeroth-order bands (L0,M0) appear  when the self-energy barrier $E_{self}$ is balanced by the site attraction energy  $E_{attr}$  (\ref {Sec:energetics1ion}). The self-energy barrier $E_{self}$  can be estimated by application of Gauss's theorem to the channel volume taking account of the near-zero radial field as \cite{Zhang:05}:
\begin {equation}
   E_{self}= \frac {1}{4\pi\varepsilon_0}\frac {(ze)^2 L}{2\varepsilon_w R^2}.
\label{equ:gauss_self}
\end {equation}
A similar approach gives us for the attraction energy
\begin {equation}
   E_{attr}= \frac {1}{4\pi\varepsilon_0}\frac {(ze) Q_f L}{\varepsilon_w R^2}
\label{equ:gauss_attr}
\end {equation}
From condition  $E_{self}=E_{attr}$ we take the result:
\begin {equation}
   Q_f(0,z)= z \cdot Q_f(0,1)=\frac {(ze)}{2}
\label{equ:gauss_qf}
\end {equation}
We can define the effective image charge for an ion of charge $ze$ as being $Q_{\rm eff}=(ze)/2$ and thus interpret (\ref {equ:gauss_qf}) as the neutralisation condition $Q_f=Q_{\rm eff}$. The multiplier (1/2) appears from the textbook formula for the electrostatic self-energy $E_{\rm self}=(1/2)e U_{rf}$, where $U_{rf}$ stands for the potential of the reaction field \cite{Cheng:05, Zhang:05}.

Thus, we may expect that next resonances will appear periodically at intervals of $ze$ (i.e.\ 1$e$ for Na$^+$, 2$e$ for Ca$^{2+}$) when  additional $Q_f$ is neutralised by the charge of an integer number of sequentially captured ions (L1,M1,...). For an ion of valence $z$ one can write:
\begin {equation}
Q_f(i,z)=Q_f(0,z)+z e \cdot i= z e(\frac{1}{2}+ i )
 \label{equ:main}
\end {equation}
where the order of the band $i$=0,1,2... is equal to the number of ions captured by the site, i.e.\ the saturated site occupancy ($i=0$ for L0 and M0, etc).

\begin{figure}[t]
\includegraphics[width=1.0\linewidth]{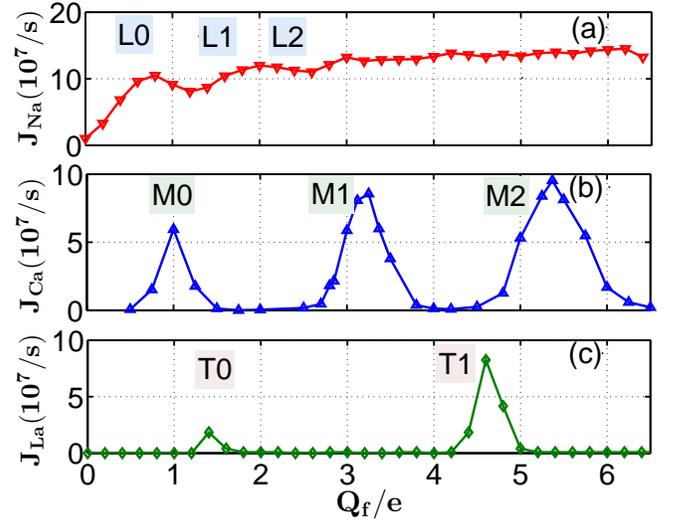}
\caption{(Color online)  BD simulations showing the pattern of conduction bands and valence selectivity for Na$^+$, Ca$^{2+}$ and La$^{3+}$ ions. (a) The conduction bands for Na$^+$ are overlapped and smooth (replotted for easier comparison from Fig.\ \ref{fig:na_bands}(a)).  (b) Ca$^{2+}$ conduction exhibits clearly resolved peaks M0, M1, M2 (c) La$^{3+}$ conduction exhibits the clearly resolved peaks T0 and T1} \label{fig:la_bands}
\end{figure}

Equation (\ref{equ:main}) provides for a separation in $Q_f$ space of the bands of ions for different valence $z$ and hence gives rise to valence selectivity in the generic model ion channel. That is, for our rigid, fixed-charge model, we will have the following sequences of conduction bands for ions of different valence (cf.\ Fig.\ \ref {fig:la_bands}) --

\begin{enumerate}

\item \underbar{Monovalent Na$^{+}$ ions}
\begin {equation}
Q_f(i,1)=0.5e({\rm L0})\rightarrow 1.5e({\rm L1})\rightarrow 2.5e({\rm L2})...
\label {equ:mono}
\end{equation}
The band positions  (\ref{equ:mono}) are also in good agreement with the energetically defined pattern from electrostatics. However, the BD-simulated bands shown in Fig.\ \ref{fig:la_bands}(a) appear at somewhat higher values of $Q_f$ (see also Sec.\ \ref{Sec:pattern}), a discrepancy that requires further investigation.

\item \underbar{Divalent Ca$^{2+}$ ions}
\begin {equation}
Q_f(i,2)=1e({\rm M0})\rightarrow3e({\rm M1})\rightarrow 5e({\rm M2})...
\label {equ:divalent}
\end{equation}
These predictions (\ref{equ:divalent})  are in a good agreement with the patterns of Ca$^2+$ conduction bands seen in both the BD-simulations (Fig.\ \ref{fig:la_bands}(b)) and the electrostatic calculations (Sec.\ \ref{Sec:energetics1ion}).

\item \underbar{Trivalent La$^{3+}$ ions}
\begin {equation}
Q_f(i,3)=1.5e({\rm T0})\rightarrow4.5e({\rm T1})...
\label {equ:trivalent}
\end{equation}
The predictions (\ref{equ:trivalent})  agree well with the pattern of La$^3+$ conduction bands seen in the BD-simulations (Fig.\ \ref{fig:la_bands}(c)). The pattern of predicted/simulated bands also agrees with the experimentally observed blockage of the Ca$^{2+}$ current by trivalent ions \cite{Lansman:90}

\end{enumerate}

It follows from (\ref {equ:main}) that the channel does not conduct when its fixed charge $Q_f$ is completely neutralised by the sum of the charges of captured ions so that the channel (or its selectivity filter) is electrically neutral. Such a {\it neutralised state} can appear only  for integer values of $Q_f/(ze)$ due to the discreteness of the charge. Neutralised  states with integer $Q_f/(ze)$ are charge-saturated and stable, whereas non-neutralised states with half-integer $Q_f/(ze)$ exhibit high conductivity and selectivity.

This neutralisation  approach is close to the space-charge competition model of the calcium channel \cite{Boda:07} and to the one-dimensional Coulomb gas theory of ionic motion inside an ion channel developed in \cite{Zhang:05, Kamenev:06}.
The pattern of bands is similar to the energy level structure of a quantum harmonic oscillator and is also reminiscent of that seen in the quantum Hall effect \cite{Stone:81}.

\section{Conclusions}\label{Sec:Conclusions}

In summary, we have carried out Brownian dynamics simulations of ionic conduction in a generic model of a channel in the calcium-sodium channel family, for different values of the negative charge at the selectivity filter $Q_f=0-6.5e$. They reveal a strictly ordered sequence of selectivity bands of increased calcium selectivity: L0=0.5$e$ (sodium selective, non-blocking channel) $\rightarrow$ M0=1$e$ (non-selective cation channel) $\rightarrow$ L1=1.5$e$ (sodium selective, blocking channel) $\rightarrow$ M1=3$e$ (calcium selective, blocking channel with AMFE, single-ion block) $\rightarrow$ M2=5$e$ (calcium selective, blocking channel with AMFE, double-ion block). Conduction bands correspond to ion-exchange phase transitions obtained analytically in \cite{Zhang:06}

Our preliminary identification of bands \cite{Kaufman:13a} has been confirmed, and completed as follows: L0 corresponds 
the eukariotic DEKA sodium channel (inner ring); M0 to the non-selective cation channel or to OmpF porin; L1 to the LNCN sodium channel and to the outer EEEE ring of eukaryotic sodium channel and to main EEEE locus of bacterial sodium channels; M1 to the L-type EEEE calcium channel; and M2 to the RyR DDDD calcium channel.

The completed identification scheme accounts for the experimentally observed mutation transformations of conductivity/selectivity between the non-selective channel, sodium channels and calcium channels. It is suggested that mutation-induced transformations appear as transitions between different rows in the identification table. The scheme provides a unified and straightforward explanation for the results of several mutation studies in the Ca$^{2+}$/Na$^+$ family of ion channels and in OmpF porin.

By consideration of optimal trajectories on potential energy surfaces, our investigations of the energetics of conduction and valence selectivity have shown explicitly, that the multi-ion conduction bands of the calcium/sodium channels arise as the result of single- and multi-ion barrier-less conduction. These resonance-like effects are more pronounced for the divalent calcium bands M0 and M1 than they are for the sodium L0 and L1 bands.


Our results confirm the crucial influence of electrostatic interactions on the conduction and Ca$^{2+}$/Na$^+$ valence selectivity of calcium and sodium ion channels, thereby resolving the celebrated selectivity {\it vs.} conductivity paradox. They have also demonstrated the surprisingly broad applicability of generic ion channel models. We speculate that they they may readily be extended to describe the permeation and selectivity properties of artificial nanopores.

\section*{Acknowledgements}
The research was supported by the Engineering and Physical Sciences Research Council UK (grant No.\ EP/G070660/1).

I.K. would like to thank Prof. B.I. Shklovskii for a helpful discussion.

\end{document}